\documentclass{article}
\usepackage{multirow}
\usepackage{amsmath}
\usepackage{amsfonts}
\usepackage{amssymb}
\usepackage{amsthm}
\usepackage{wasysym}
\usepackage{tikz}
\usepackage{calc}
\usepackage{hyperref}
\usepackage[ruled,vlined]{algorithm2e}
\usepackage{cleveref}
\usepackage{textcomp}
\usetikzlibrary{decorations.pathreplacing,automata,arrows,patterns,shapes,chains}
\usepackage{graphicx}
\usepackage{xcolor}
\usepackage{colortbl}
\usepackage[left=3cm,right=3cm,top=3cm,bottom=3cm]{geometry}
\theoremstyle{definition}
\newtheorem{definicion}{Definition}
\newtheorem{ejemplo}{Example}
\newtheorem{observacion}{Remark}

\theoremstyle{plain}
\newtheorem{teorema}{Theorem}
\newtheorem*{thm}{Theorem}
\newtheorem{lema}{Lemma}
\newtheorem{proposicion}{Proposition}
\newtheorem{corolario}{Corollary}

\newenvironment{demostracion}[1][.]{\paragraph{Proof\ifx#1..\else ~of #1.\fi}}{\hfill $\Box$\par}
\renewcommand{\;}{;\\}

\newenvironment{algoritmo}[2][]{\begin{algorithm}[htb]\caption{#2}\label{#1}}{\end{algorithm}}

\newcommand{\N}{\mathbb{N}}
\newcommand{\FVS}{\operatorname{FVS}}
\newcommand{\PFVS}{\operatorname{PFVS}}
\newcommand{\Set}[1]{\left\{#1\right\}}
\newcommand{\B}{\Set{0,1}}
\newcommand{\bn}[1][f]{#1: \B^n \to \B^n}
\newcommand{\funca}{\mathit{fa}}
\newcommand{\Abfp}[1][.]{\ifx#1.BasicFixedPoint algorithm \else BasicFixedPoint \fi}
\newcommand{\Afp}[1][.]{\ifx#1.FixedPoint algorithm \else FixedPoint \fi}
\newcommand{\Apfvs}[1][.]{\ifx#1.PFVS algorithm \else PFVS \fi}

\newcommand{\anc}{\operatorname{Anc}}
\newcommand{\dec}{\operatorname{Dec}}
\newcommand{\bef}{\operatorname{Bef}}
\newcommand{\aft}{\operatorname{Aft}}

\newcommand{\jf}{I_k}
\newcommand{\fjf}{f^{\jf}}
\newcommand{\highli}[1]{\cellcolor{blue!25}#1}
\newlength{\rnodo}
\tikzstyle{vertex}=[outer sep=1,inner sep=0,circle,draw,minimum size=\rnodo,fill=white,font=\scriptsize]
\tikzstyle{labels}=[inner sep=0pt,font=\tiny,auto,circle]
\tikzstyle{arcos}=[-latex, thick]
\tikzstyle{dark}=[fill=black, text=white]
\tikzstyle{semi}=[fill=black!50, text=white]
\tikzstyle{light}=[fill=black!10, text=black]

\colorlet{verde}{green!70!black}
\colorlet{rojo}{red}

\newcommand{\Bucle}[4][]{\draw[arcos, #1](#2.#3+10)..controls +(#3+20:\rnodo) and +(#3-20:\rnodo).. node[labels]{#4}(#2.#3-10)}

\newcommand{\flecha}[1]{\begin{tikzpicture}[baseline=(e)]\coordinate (e) at (0,-2pt);\draw[arcos,#1](0,0) --(0.7cm,0);\end{tikzpicture}}
\begin{document}

\title{Finding the fixed points of a Boolean network from a positive feedback vertex set}

\author{
Julio Aracena\footnote{CI2MA and Departamento de Ingenier\'ia Matem\'atica, Universidad de Concepci\'on, Chile.},
Luis Cabrera-Crot \thanks{Corresponding author} \footnote{Departamento de Ing. Inform\'atica y Cs. de la Computaci\'on, Universidad de Concepci\'on, Chile.},
Lilian Salinas\footnote{Departamento de Ing. Inform\'atica y Cs. de la Computaci\'on and CI2MA, Universidad de Concepci\'on, Chile.}
}
%

\maketitle

\begin{abstract}
In the modeling of biological systems by Boolean networks a key problem is finding the set of fixed points of a given network. Some constructed algorithms consider certain structural properties of the interaction graph like those proposed by Akutsu et al. in \cite{akutsu1998system,zhang2007algorithms} which consider a feedback vertex set of the graph. However, these methods do not take into account the type of action (activation, inhibition) between its components.

In this paper we propose a new algorithm for finding the set of fixed points of a Boolean network, based on a positive feedback vertex set $P$ of its interaction graph and which works, by applying a sequential update schedule, in time $O(2^{|P|} \cdot n^2)$, where $n$ is the number of  components. The theoretical foundation of this algorithm is due a nice characterization, that we give, of the dynamical behavior of the Boolean networks without positive cycles and with a fixed point.

An executable file of \Afp  made in Java and some examples of input files are available at: \href{http://www.inf.udec.cl/~lilian/FPCollector/}{\url{www.inf.udec.cl/~lilian/FPCollector/}}
\end{abstract}

\section{Introduction}

A Boolean network is a system of $n$ interacting Boolean variables, which evolve, in a discrete time, according to a evolution rule and to a predefined updating scheme. Boolean networks have many applications, including computer science, in particular circuit theory,  and social systems. In particular, from the seminal works of S. Kauffman \cite{Kau69,Kau93} and R. Thomas \cite{Thom73,Thom90}, they are extensively used as models of gene networks. In this context, the fixed points of a network are associated to distinct types of cells defined by patterns of gene activity \cite{aracena2006regulatory,huang1999gene,Kau69}.
 The problem of finding the steady states or fixed points of a Boolean network is a difficult task. In \cite{akutsu1998identification}, Akutsu et al. show that deciding if a network has a fixed point is NP-complete, consequently, counting how many fixed points a Boolean network has is \#P-complete. For this reason, in the literature, several strategies have been proposed to find fixed points in Boolean networks. Some of them are:
 Reduction method \cite{veliz2011reduction,veliz2015dimension,zanudo2013effective}, Representation as polynomial functions \cite{hinkelmann2011adam,zou2013algorithm}, SAT-based methods \cite{rolf2006improved,akutsu2011determining,devloo2003identification,dubrova2011sat,melkman2010determining,tamura2009algorithms,tamura2009detecting}, Methods based on Integer Programming \cite{akutsu2009integer}, Strategic Sampling \cite{zhang2007algorithms} and Methods based on Minimal Feedback Vertex Sets \cite{akutsu1998system,zhang2007algorithms}. For a more detailed description of these and other methods see \cite{veliz2014steady,he2016algorithm} and references therein.

The structure of a Boolean network is often represented by a digraph, called interaction graph, where vertices are network components, and there is an arc from one component to another when the evolution of the latter depends on the evolution of the former. In some cases, these arcs can have associated a positive or a negative sign if the action associated is of type activation or inhibition respectively. 

Some of the algorithms for finding the fixed points of a Boolean network consider certain characteristics of its interaction graph. In particular,  Akutsu et al. in \cite{akutsu1998system,zhang2007algorithms} proposed a method consisting in fixing the values of the vertices of a minimal feedback vertex set $F$ and propagating them to the others. In this way the problem of finding the fixed points of a Boolean network of $n$ components is reduced to check $2^{|F|}$ state configurations instead of $2^n$, where usually  the size of $F$ is less than $n$. However, this method does not consider the sign of the arcs of the interaction graph.

On the other hand, to our knowledge it does not exist a method to find the fixed points of a Boolean network that take advantage of the fact they are invariant to the update schedule.

In this paper we propose a new strategy to find the fixed points of a Boolean network which consists in fixing the local states of a  positive feedback vertex set $P$ of the interaction graph and updating the other local states according to a sequential update schedule constructed from $P$. In this way we reduce the problem of finding the fixed points of a Boolean network of $n$ components to check $2^{|P|}$ state configurations, where $P$ can be smaller than a feedback vertex set of the network. Thus, the proposed algorithm is quadratic in the size of the network and exponential in the size of $P$. Consequently, this method can be very efficient, with respect to other methods, in Boolean networks with a small positive feedback vertex set, as for example ones with few positive cycles. A particular case of this latter are the strong inhibition Boolean networks \cite{he2016algorithm}. 

For the construction of the proposed algorithm we proved first that the dynamical behavior of a Boolean network without positive cycles and with a fixed point is similar to an acyclic Boolean network, i.e. it quickly converges to the fixed point from any initial configuration.

\section{Definition and notation}
A \emph{Boolean network} (BN) with $n$ components is a discrete dynamical system usually defined by a \emph{global transition function}:
\[ 
    \bn, \  x \to f(x) = (f_1(x),\dots,f_n(x)), 
\]
where each function $f_i: \B^n \to \B$ associated to the component $i$ is called \emph{local activation function}.

Any vector $x=(x_1, \dots, x_n) \in \B^n$ is called a \emph{state} of the network $f$ with \emph{local state} $x_i$ on each component $i$. 
The \emph{dynamics} of $f$ is given by its application on any state of the network. A \emph{steady state or fixed point} of $f$ is any state $x \in \B^n$ such that $f(x) = x$.

We define the \emph{interaction graph} of a BN $f$ with $n$ components, denoted $G(f)=(V,A)$, as the  signed directed graph with a sign, +1 or -1, attached to each arc, where $V$ is a finite set of elements called vertices or nodes indexed in the set $[n]:=\Set{1,2,\dots,n}$ and $A$ is a collection of ordered pairs of vertices of $V$, called arcs, such that  $\forall u,v \in [n]:$

\begin{itemize}
    \item $(u,v) \in A$, with sign $+1$, if $\exists x \in \B^n$, $x_u=0$, \mbox{$f_v(x) < f_v(x + e_u)$},
    \item$(u,v) \in A$, with sign $-1$, if $\exists x \in \B^n$, $x_u=0$, \mbox{$f_v(x) > f_v(x + e_u)$},
\end{itemize}

where \mbox{$e_u \in \B^n$} denotes the binary vector with all entries equal to 0, except for entry $u$, which equals 1. The vertex set of $G(f)$ is referred to as $V(G(f))$ and its arc set as $A(G(f))$. Note that $G(f)$ can have both a positive and a negative arc from one vertex to another. If $G(f)$ does not have  multiple arcs from one vertex to another, then we say that $f$ is a regulatory Boolean network (RBN).

A \emph{walk} from a vertex $v_0$ to a vertex $v_l$ in the interaction graph $G(f)$ is a sequence of vertices and arcs $W = v_0,a_0,v_1,\dots,a_{l-1},v_l$ of $G(f)$ such that $\forall i\in\Set{0,\dots,l-1}$, $a_i=(v_i,v_{i+1})\in A(G(f))$. $A(W) = \Set{a_i : i \in \Set{0,\dots, l-1}}$ is the set of arcs of $W$.  A \emph{path} is a walk without repetition of vertices (except eventually the extreme ones). A \emph{circuit} is a walk without repetition of arcs and closed (i.e. its extreme vertices are equal). A \emph{cycle} is a closed path.

Let $u \in V(G(f))$, we denote the in-degree and out-degree of $u$ by
\mbox{$d^-(u) = |\Set{v \in V(G(f)): (v,u) \in A(G(f))}|$}  and \mbox{$d^+(u) = |\Set{v \in V(G(f)): (u,v) \in A(G(f))}|$}, respectively. If $d^-(u)=0$, we say that $u$ is a \emph{source vertex}. The degree of $u$ is $d(u)=d^-(u)+d^+(u)$. We refer \cite{bang2008digraphs} for other basic definitions in digraphs.

The sign of a walk (path, circuit or cycle) is the product of the signs of its arcs. We say that a cycle is a positive cycle, if the sign of the cycle is $+1$, otherwise, we say that is a negative cycle. See Example~\ref{ex:signed-digraph}.

A vertex set $F \subseteq V(G(f))$ is a \emph{feedback vertex set} (FVS) of $G(f)$ if $G(f)-F$ is acyclic. $F$ is said to be a \emph{minimal feedback vertex set} of $G(f)$ if $F$ is a FVS of $G(f)$ and there is no other FVS $F'$ such that $F' \subsetneq F$. $F$ is said to be a \emph{minimum feedback vertex set} of $G(f)$ if $F$ is a FVS of $G(f)$ and there is no other FVS set $F'$ such that $|F'| < |F|$.

We denote $\tau(G(f)) = \min\Set{|F|: F \text{ is a FVS of }G(f)}$ the \emph{transversal number} of $G(f)$. 

$P \subseteq V(G(f))$ is a \emph{positive feedback vertex set} (PFVS) of the signed interaction graph  $G(f)$ if $G(f)-P$ is a digraph without positive cycles. The minimum cardinality of a PFVS is denoted by $\tau^{+}(G(f))$ and called the \emph{positive transversal number} of $G(f)$.

When there is no confusion, for simplicity we denote $\tau (G(f))$ by  $\tau$ and $\tau^+ (G(f))$ by  $\tau^+$.

\begin{observacion}\label{tauG-geq-tauplusG}
Since a FVS is a particular PFVS,
\[ 
    \tau^{+} \leq \tau. 
\]
\end{observacion}

\begin{ejemplo}\label{ex:signed-digraph}
An example of a BN $f$ and its interaction graph $G(f)$ is shown in \Cref{fig:signed-graph}. Positive arcs are represented by \flecha{verde} and negative arcs by \flecha{rojo}.
Examples of positive cycles are: $C_1=4,a_{12},5,a_{11},4$ and $C_2=3,a_4,2,a_2,1,a_8,3$, and  examples of negative cycles are: $C_3=3,a_6,2,a_2,1,a_8,3$ and $C_4=1,a_3,2,a_2,1$. Dark vertices are a PFVS of $G(f)$.

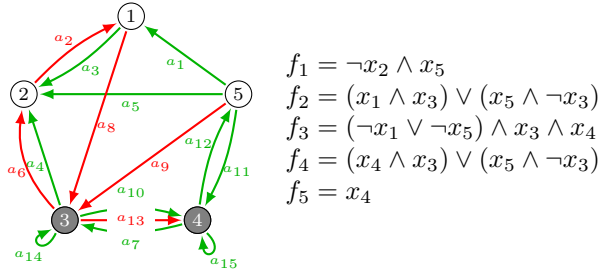
\begin{figure}[htb]
	\centering
	\begin{tikzpicture}
	\foreach \x in {1,...,5}{
		\node[vertex] (\x) at (18+72*\x:1.5\unitlength) {\x};
	}
		\node[vertex,semi] (4) at (306:1.5\unitlength) {4};
		\node[vertex,semi] (3) at (234:1.5\unitlength) {3};
	\node[right] (l) at (1.75\unitlength,0) {%
	$\begin{array}{l}
		f_1 = \lnot x_2 \land x_5\\
		f_2 = (x_1 \land x_3) \lor (x_5 \land \lnot x_3)\\
		f_3 = (\lnot x_1 \lor \lnot x_5) \land x_3 \land x_4\\
		f_4 = (x_4 \land x_3) \lor (x_5 \land \lnot x_3)\\
		f_5 = x_4
	\end{array}$};
	\Bucle[verde]{3}{225}{$a_{14}$};
	\Bucle[verde]{4}{296}{$a_{15}$};
	\path[arcos,verde]
		(5) edge node[labels]{$a_1$} (1)
		(1) edge[bend left =10] node[labels]{$a_3$} (2)
		(3) edge node[labels]{$a_4$} (2)
		(5) edge node[labels]{$a_5$} (2)
		(4) edge[bend left =15] node[labels]{$a_7$} (3)
		(3) edge[bend left =15] node[labels]{$a_{10}$} (4)
		(5) edge[bend left =10] node[labels]{$a_{11}$} (4)
		(4) edge[bend left =10] node[labels]{$a_{12}$} (5)
	;
	\path[arcos,rojo]
		(2) edge[bend left =10] node[labels]{$a_2$} (1)
		(3) edge[bend left = 25] node[labels]{$a_6$} (2)
		(1) edge node[labels]{$a_8$} (3)
		(5) edge node[labels]{$a_9$} (3)
		(3) edge node[font=\tiny,fill=white]{$a_{13}$} (4)
	;
	\end{tikzpicture}
	\caption{Example of a BN and its interaction graph.}
	\label{fig:signed-graph}
\end{figure}

\end{ejemplo}

In the sequel, we will refer to the signed cycles of a BN $f$ as the cycles of $G(f)$.

Given $\bn$ a BN and $u \in [n]$, the \emph{partial evaluation} of $f$ in the local function $f_u$ is the function $\bn[f^u]$ such that:
\[
    \forall x \in \B^n, \ f^u(x) = (x_1,\dots,x_{u-1},f_u(x),x_{u+1},\dots,x_n)
\]

We define a \emph{sequential schedule}  $\pi=(\pi_1,\pi_2,\dots,\pi_n)$ as a permutation of $[n]$. The dynamics of a BN $\bn$ updated according to $\pi$ is defined by:
\[
 \forall x \in \B^n,  \  f^\pi(x) = f^{\pi_n}\circ f^{\pi_{n-1}}\circ\dots\circ f^{\pi_1}(x).
\]

To avoid confusion, we use a special notation for self-compositions of $f$: we denote by $f
^{\langle0\rangle}$ the identity on $\B^n$ and for $k \geq 1$ we set $f^{\langle k \rangle}= f \circ f^{\langle k-1 \rangle}$. The concatenation of $k$ times the sequential schedule $\pi$ is denoted $\pi^k$. In this way $f^{\pi^k}=(f^{\pi})^{\langle k\rangle}$.

\section{Boolean networks without positive cycles}
The relationship between the fixed points and the positive and negative cycles in Boolean networks has been greatly studied \cite{Aracena08,aracena2017fixed,ARSsidma2017,remy2008graphic,richard2018fixed,richard2019positive}. In particular, in  \cite{Aracena08, remy2008graphic} was proved that every Boolean network without positive cycles has at most one fixed point and, in addition, if  the network has an initial non-trivial strong component, then it has no fixed points.
On the other hand, F.  Robert studied the Boolean networks without cycles and  proved in \cite{Robert86,robert1995systemes} that these networks have a simple dynamical behavior with a unique attractor, which is a fixed point, and where all initial states quickly converge to it.  More precisely, he proved  the following theorem. 
\begin{thm}[\bf F. Robert]
Let $\bn$ be a BN such that $G(f)$ is acyclic. Then,
\begin{enumerate}
    \item $f$ has a unique fixed point $y \in \B^n$.
    \item $\forall x \in \B^n, \ f^{\langle n\rangle}(x)=y$.
    \item $\exists \pi$ a sequential  schedule such that $\forall x \in \B^n, \ f^{\pi}(x)=y$.
\end{enumerate}
\end{thm}
In this paper we show that whether a BN without positive cycles has a fixed point, then its dynamical behavior is like a BN without cycles. More precisely, we prove the following theorem. 
\begin{teorema}\label{teo:maestra}
Let $\bn$ be a BN without positive cycles. Then, the following propositions are equivalents: 
\begin{enumerate}
    \item $f$ has a unique fixed point $y \in \B^n$.
    \item $\forall x \in \B^n, \ f^{\langle n\rangle}(x)=y$.
    \item $\exists \pi$ a sequential  schedule such that $\forall x \in \B^n, \ f^{\pi}(x)=y$.
    \end{enumerate}
\end{teorema}

To proof this result, we need to introduce some definitions and previous results.

\begin{definicion}\label{def:setI}
Let $\bn$ be a Boolean network, we denote the following sets (eventually empty):
\[
I_1(f) = \Set{v \in [n]: \exists c \in \B, \forall x \in \B^n, f_v(x) = c}.
\]

In other words, $v \in I_1(f)$ if $f_v$ is  constant. For all $v \in I_1(f)$, we denote $c_v$ to the constant value of $f_v(x)$.

Recursively, we define $\forall k \in \N, k \geq 2, I_k(f)$ in the following way:
\[
I_k(f) = \{v \in [n] : \exists c \in \B, \forall x \in \B^n,  f_v(c_u : u \in I_{k-1}(f); x_u : u \notin I_{k-1}(f)) = c\}.
\]

Similarly, $v \in I_k(f)$ if the local activation function associated to the component $v$ with fixed value $c_u$ on input $u \in I_{k-1}(f)$ is a constant function. $\forall v \in I_k(f)$, we denote $c_v$ to the constant value of \mbox{$f_v(c_u : u \in I_{k-1}(f); x_u : u \notin I_{k-1}(f))$}.
\end{definicion}

From here, if $v \in I_k(f)$, then:
\[ \forall t \geq k, \forall x \in \B^n, f^{\langle t\rangle}_v(x) = c_v.\]
In this case, we say that the component $v$ is \emph{fixed at iteration $k$ of $f$ with value $c_v$.}
\begin{proposicion}\label{prop:Ione}
The sets $I_k(f)$ have the following properties:
\begin{enumerate}
\item $\forall k \geq 1, I_{k}(f) \subseteq I_{k+1}(f).$
\item If $\exists k \in \N, I_k(f)=I_{k+1}(f)$, then $\forall l \in \N, I_k(f)=I_{k+l}(f)$.
\item If $I_1(f)=\emptyset$, then $\forall k \in \N, I_k(f)=\emptyset$.
\end{enumerate}
\end{proposicion}

\begin{demostracion}
\begin{enumerate}

\item By induction on $k$. First, we prove that $I_1(f) \subseteq I_2(f)$. By definition, $v \in I_1(f)$ if $f_v$ is a constant function. On the other hand, $v \in I_2(f)$ if $f_v$ with fixed value $c_u$ on each input $u \in I_1(f)$ is a constant function. Therefore, $I_1(f) \subseteq I_2(f)$. 

Let us suppose that $I_k(f) \subseteq I_{k+1}(f)$. 
Now, we prove that $I_{k+1}(f) \subseteq I_{k+2}(f)$. Analogously, $v \in I_{k+1}(f)$ if $f_v$ with fixed value $c_u$ on each input $u \in I_k(f)$ is a constant function. On the other hand, $v \in I_{k+2}(f)$ if $f_v$ with fixed value $c_u$ on each input $u \in I_{k+1}(f)$ is a constant function. Since $I_k(f) \subseteq I_{k+1}(f)$, the inclusion $I_{k+1}(f) \subseteq I_{k+2}(f)$ is direct.

\item To prove this, we prove that if $I_k(f) = I_{k+1}(f)$ then $I_{k+1}(f) = I_{k+2}(f)$.

Let us suppose that $I_k(f) = I_{k+1}(f)$. If $I_{k+2}(f)= \emptyset$, by Property 1, $I_{k+1}(f)=\emptyset$. So, let $v$ be a component in $I_{k+2}(f)$, by definition:
\[\exists c \in \B, \forall x \in \B^n, f_v(c_u : u \in I_{k+1}(f); x_u : u \notin I_{k+1}(f)) = c.\]
Since $I_k(f) = I_{k+1}(f)$,
\[\forall x \in \B^n, f_v(c_u : u \in I_k(f); x_u : u \notin I_k(f)) = c.\]
Therefore, $v \in I_{k+1}(f)$. For this reason, $I_{k+2}(f) \subseteq I_{k+1}(f)$ and since $I_{k+1}(f) \subseteq I_{k+2}(f)$, then $I_{k+1}(f) = I_{k+2}(f)$.

\item First, we prove that if $I_1(f) = \emptyset$, then $I_2(f) = \emptyset$.

By contradiction, let us suppose that $I_1(f) = \emptyset$ and $I_2(f) \neq \emptyset$. Let $v$ be a component in $I_2(f)$, then:
\[\exists c \in \B, \forall x \in \B^n, f_v(c_u : u \in I_1(f); x_u : u \notin I_1(f)) = f_v(x) = c.\]
Therefore, by definition, $v \in I_1(f)$ which is a contradiction.

Since $I_1(f) = I_2(f) = \emptyset$, by Property 2, $\forall k \in \N, I_k(f) = \emptyset$.
\end{enumerate}
\end{demostracion}

\begin{definicion}\label{def:fr}
Let $\bn$ be a Boolean network. We say that $f$ is an \emph{irreducible network after $k$ iterations} or \emph{$k$-irreducible network} if $I_{k}(f) = I_{k+1}(f)$. If $I_1(f) = \emptyset$, we say that $f$ is an \emph{irreducible network}.
\end{definicion}

\begin{definicion}\label{def:fi}
Let $\bn$ be a Boolean network and $k \in \mathbb{N}$. We define \mbox{$\bn[\fjf]$} the \emph{Boolean network associated to $f$ and $\jf(f)$} by:
\[\forall x \in \B^n, \fjf(x) = f(c_u: u \in \jf(f); x_u: u \notin \jf(f)).\]
\end{definicion}

\begin{observacion}
Note that if $f$ is an irreducible network, then $f^{I_k}(x) = f(x)$.	
\end{observacion}

\begin{ejemplo}
\Cref{fig:fi} shows an example of a Boolean network $f$ and the network $f^{I_3}$, according to \Cref{def:fi}. 
Since $f_4$ and $f_5$ are the only constant functions, $I_1(f) = \Set{4,5}$ (with $c_4 = 1$ and $c_5 = 0$). On the other hand, \mbox{$f_1(x_1,x_2,x_3,1,0,x_6,x_7)= \overline{x_1} \vee 1 \vee 0 = 1$}, hence $1 \in I_2(f)$. Moreover, since there is no other component that satisfies the same condition, $I_2(f) = \Set{1,4,5}$.

Analogously, $I_3(f) = \Set{1,4,5,6,7}$. In a new iteration, $I_4(f) = I_3(f)$, so $f$ is $3$-irreducible.
\begin{figure}[!htb]
\centering
\begin{tikzpicture}[baseline=(7)]
\coordinate (c1) at (120:\unitlength);
\coordinate (c2) at (0:\unitlength);
\coordinate (c3) at (300:\unitlength);
\coordinate (c4) at (240:\unitlength);
\coordinate (c5) at (180:\unitlength);
\coordinate (c6) at (60:\unitlength);
\coordinate (c7) at (0,0);
\coordinate (t1) at (0:1.5\unitlength);
\coordinate (t2) at (0:2.5\unitlength);
\foreach \x in {1,...,7}{
    \node[vertex] (\x) at (c\x) {\x};
}
\node[right] (l) at (0:1.5\unitlength) {\footnotesize $\begin{array}{l}
f_1 = \overline{x_1} \vee x_4 \vee x_5\\
f_2 = \overline{x_2} \wedge x_3 \wedge x_6\\
f_3 = \overline{x_3} \wedge x_4\\
f_4 = 1\\
f_5 = 0\\
f_6 = x_1 \vee x_3 \vee x_7\\
f_7 = x_1 \wedge x_4\\
\end{array}$};
\node[below] (l2) at (1.5\unitlength,-1.5\unitlength) {\footnotesize $\begin{array}{l}
I_1(f) = \Set{4,5} \text{ with } c_4 = 1 \text{ and } c_5 = 0\\
I_2(f) = \Set{1,4,5} \text{ with } c_1 = 1\\
I_3(f) = \Set{1,4,5,6,7} \text{ with } c_6 = 1 \text{ and } c_7 = 1
\end{array}$};
\Bucle{1}{120}{};
\Bucle{2}{0}{};
\Bucle{3}{300}{};
\path[arcos,bend left=10]
	(1) edge (6)
	(1) edge (7)
	(3) edge[bend right=10] (2)
	(3) edge (6)
	(4) edge (1)
	(4) edge[bend right=10] (3)
	(4) edge (7)
	(5) edge (1)
	(6) edge (2)
	(7) edge (6)
;
\end{tikzpicture}
\hfill
\begin{tikzpicture}[baseline=(7)]
\node (0) at (-1.2\unitlength,\unitlength) {$f^{I_3}:$};
\foreach \x in {1,...,7}{
    \node[vertex] (\x) at (c\x) {\x};
}
\node[right] (l) at (0:1.5\unitlength) {\footnotesize $\begin{array}{ll}
f^{I_3}_1(x) = \overline{1} \vee 1 \vee 0 &= 1\\
f^{I_3}_2(x) = \overline{x_2} \wedge x_3 \wedge 1&= \overline{x_2} \wedge x_3  \\
f^{I_3}_3(x) = \overline{x_3} \wedge 1 &= \overline{x_3}\\
f^{I_3}_4(x) = 1 &= 1\\
f^{I_3}_5(x) = 0 &= 0\\
f^{I_3}_6(x) = 1 \vee x_3 \vee x_7 &= 1\\
f^{I_3}_7(x) = 1 \wedge 1 &= 1\\
\end{array}$};
\Bucle{2}{0}{};
\Bucle{3}{300}{};
\path[arcos,bend right=10] (3) edge (2);
\end{tikzpicture}
\caption{Example of Boolean networks $f$ and $f^{I_3}$.}\label{fig:fi}
\end{figure}
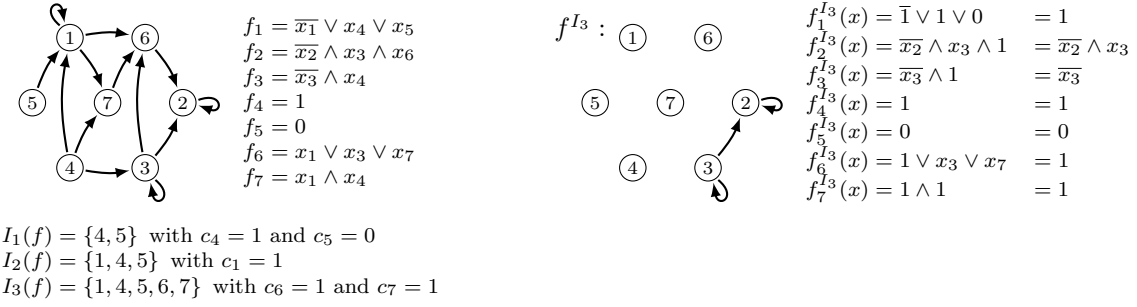
\end{ejemplo}

\begin{observacion}\label{rem:fi}
It is important to remark that $\forall k \in \mathbb{N}$ and for all \mbox{$\bn$}:
\begin{enumerate}
\item $V(G(\fjf)) = V(G(f))$.
\item $A(G(\fjf)) \subseteq A(G(f))$.
\item If $(u,v) \in A(G(\fjf))$, then the sign of $(u,v)$ is the same in $G(f)$ and $G(\fjf)$.
\item $y$ is a fixed point of $f$ if and only if $y$ is a fixed point of $\fjf$ and $\forall v \in I_k(f), c_v = y_v$.
\end{enumerate}
\end{observacion}

\begin{observacion}\label{rem:nscc}
Let $\bn$ be a Boolean network such that $I_k(f) = I_{k+1}(f) \neq \emptyset$. Then, $G(\fjf)$ has the following properties:
\begin{enumerate}
\item $\forall u \in I_k(f), \ d^+_{G(\fjf)}(u) = 0$.
\item If there exists a non-trivial strong component, then there is an initial non-trivial strong component.
\end{enumerate}
\end{observacion}


In \cite{Aracena08,remy2008graphic} was proved the following theorem, known as the Thomas's first rule:
\begin{thm}[\bf Thomas's first rule]
Let $\bn$ a BN without positive cycles. Then, $f$ has at most one fixed point. 
\end{thm}
A direct corollary from previous theorem is that if $f$ is a BN without positive cycles  such that has an initial non-trivial strong component, then $f$ has no fixed points \cite{Aracena08}. 

From  theorem Thomas's first rule we can state the following result.
\begin{lema}\label{lema:inclusionI}
Let $\bn$ be a RBN without positive cycles. $f$ has a unique fixed point if and only if there exists $k\in\Set{1,\dots,n}$ such that $ \emptyset\neq I_1(f)\subset I_2(f) \subset \dots \subset I_{k}(f)=[n]$.
\end{lema}
\begin{demostracion}
$(\Rightarrow)$ Let us suppose that $f$ has a unique fixed point $y \in \B^n$, we prove that $\forall k \in \Set{1,\dots,n}, |I_k| \geq k$.
Since $f$ does not have any positive cycle, by theorem Thomas's first rule all initial strong components of $G(f)$ are trivial, hence there exists a vertex $v_1 \in [n]$ whose local activation function is constant. Then, $v_1 \in I_1(f)$ and $|I_1(f)| \geq 1$.

By contradiction, let us suppose that $\exists l \in \Set{2,\dots, n}$ such that $|I_l|< l$. So, by Property 1 of \Cref{prop:Ione}:
\[ 
\exists k \leq (n-1), I_k(f) = I_{k+1}(f) \neq [n].
\]

In this case, $\forall v \notin I_k(f), \exists u \notin I_k(f)$, such that $f_v$ depends on $x_u$, \mbox{i.e., $\forall v \notin I_k(f), \exists u \notin I_k(f), (u,v) \in A(\fjf)$}.

 Then, the in-degree of all vertices in $[n]\setminus I_k(f) $ is greater than 0, hence there exists a non-trivial strong component of $G(\fjf)$ induced by the vertices in $[n]\setminus I_k(f) $. Since $\forall u \in I_k(f), \ d^+_{G(\fjf)}(u) = 0$, there exists an initial non-trivial strong component in $G(\fjf)$ and thus $\fjf$ has no fixed points. Hence, by \Cref{rem:fi}.4) $f$ has no fixed points, which is a contradiction. 

Therefore, if $k \leq (n-1)$, then $I_k(f) \subsetneq I_{k+1}(f)$ or $I_k(f) = [n]$.

$(\Leftarrow)$ If $I_n(f)=[n]$, $\forall v\in[n], \forall t\geq n, \forall x\in\B^n, f_v^{\langle t\rangle}(x) = c_v$.  Then, $f$ has a fixed point and, because $f$ does not have any positive cycles, this fixed point is unique.

\end{demostracion}
Now we give the proof of \Cref{teo:maestra}.
\begin{demostracion}[\Cref{teo:maestra}] 
The proof of $(2)\Rightarrow (1)$ and $(3)\Rightarrow (1)$ are straightforward, so we prove that $(1)\Rightarrow (2)$ and $(1) \Rightarrow (3)$.  

Both proofs are very similar, so we give the proof for $(1) \Rightarrow (2)$.  

First, notice that $\Set{I_1,I_2\setminus I_1, \dots,I_k\setminus I_{k-1}}$ is a partition of $[n]$, we define an order of the vertices $\pi=(\pi_1,\dots,\pi_n)$ such that: 
\[
\pi_i\in (I_j\setminus I_{j-1}) \ \land  \ \pi_{i'}\in (I_{j'}\setminus I_{j'-1}) \ \land \ j<j'\Longrightarrow i<i'.
\]

Now we prove by induction that $\forall i\in\Set{1\dots,k},\forall t\geq k, \ f_{\pi_i}^{\langle t\rangle}(x) = y_{\pi_i}$.

If $k=1$, $\pi_1\in I_1(f)$, then $f_{\pi_i}$ is a constant function and therefore $\forall t\geq 1$,  $f_{\pi_1}^{\langle t\rangle}(x) = y_{\pi_1}$.

Let us suppose that $\forall i\in\Set{1\dots,k},\forall t\geq k, \ f_{\pi_i}^{\langle t\rangle}t(x) = y_{\pi_i}$. 

Now we need to prove that $\forall t\geq k+1, \ f_{\pi_{k+1}}^{\langle t\rangle}(x) = y_{\pi_{k+1}}$.
\[
f_{\pi_{k+1}}^{\langle t\rangle}(x) =f_{\pi_{k+1}}(f^{\langle t-1\rangle}(x)).
\]
Since $t\geq k+1$, $t-1\geq k$, then by hypothesis of induction:
\[
f_{\pi_{k+1}}^{\langle t\rangle}(x) =f_{\pi_{k+1}}(y_{\pi_{i}}: i\leq k; \tilde x_{\pi_i}: i>k).
\]
By definition of $\pi$, if $\pi_{k+1}\in I_j\setminus I_{j-1}$ then $I_{j-1}\subseteq \Set{\pi_1,\dots,\pi_k}$, therefore, by definition of $I_j$:
\[
f_{\pi_{k+1}}^{\langle t\rangle}(x) =y_{\pi_{k+1}}.
\]

The proof for $(1)\Rightarrow (3)$ works the same way to prove that $\forall i\in\Set{1,\dots,k}$, $f_{\pi_i}^{(\pi_1,\dots,\pi_k)}(x)=f_{\pi_i}((f^{\pi_{k-1}}\circ \dots\circ f^{\pi_1})(x))=y_{\pi_i}$.
\end{demostracion}

A difference between BNs with acyclic interaction graphs and BNs without positive cycles is that, in the first case, the sequential schedule referred in F. Robert's theorem depends only on the interaction graph and not on the function. This latter is not true in the case of BNs without positive cycles as shown in  \Cref{exa:no_positive_cycles}.

\begin{ejemplo}\label{exa:no_positive_cycles} Let $f: \B^3\to \B^3$ be a BN with interaction graph shown in \Cref{fig:no_positive_cycles} and $\pi=(1,2,3)$ a sequential schedule. If we choose the functions:
 \[
    f_1(x)= 1, \qquad f_2(x) = \lnot x_1 \land x_3, \qquad f_3(x) = x_1 \land \lnot x_2,
\]
then $y=(1,0,1)$ is a fixed point of $f$ and $\forall x\in \B^3, f^{\pi}(x)=y.$
On the contrary, if we choose the functions:
\[
    f_1(x)= 1, \qquad f_2(x) = \lnot x_1 \lor x_3, \qquad f_3(x) = x_1 \lor \lnot x_2,
\]
$y'=(1,1,1)$ is a fixed point of $f$, but $f^{\pi}(1,1,0)=(1,0,1)\neq y'.$ In this case, if we choose $\pi'=(1,3,2)$, then $\forall x\in \B^3, f^{\pi'}(x)=y'$.
\begin{figure}[htp]
\centering
    \setlength{\unitlength}{1cm}
    \begin{tikzpicture}
    \node[vertex] (1) at (0,0) {1};
    \node[vertex] (2) at (\unitlength,0) {2};
    \node[vertex] (3) at (60:\unitlength) {3};
    \path[arcos,verde]
    (1) edge (3)
    (3) edge[bend left=10] (2)
    ;
    \path[arcos,rojo]
    (1) edge (2)
    (2) edge[bend left=10] (3)
    ;
    \end{tikzpicture}
    \caption{Interaction graph of a network without positive cycles.}\label{fig:no_positive_cycles}
\end{figure}
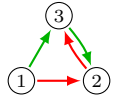
\end{ejemplo}
On the other hand, as consequence of \Cref{teo:maestra} the following corollary shows us that the problem of the existence of a fixed point in a BN without positive cycles can be solved in polynomial time.
\begin{corolario}\label{cor:linear}
Determining whether a BN $\bn$ without positive cycles has a fixed point can be done through $n+1$ applications of $f$ on any state of the network. 
\end{corolario}
\begin{demostracion}
If $y$ is a fixed point of $f$, then $\forall x \in \B^n, f^{\langle n\rangle}(x) = y$. Hence, it suffices to check for some state $x$ (for example, $x=\vec{0}:=(0,0,\dots,0)$), if $f^{\langle n\rangle}(x)$ is a fixed point of the network. In other words, if $f^{\langle n\rangle}(x)=f^{\langle n+1\rangle}(x)$ , then $f^{\langle n\rangle}(x)$ is a fixed point of $f$. Otherwise, $f$ has no fixed points.
\end{demostracion}


\section{Methods}
\subsection{Algorithm to detect the fixed points in Boolean networks from a PFVS}

In \cite{Aracena08} was proved that for every RBN $f$ there is an injection between its fixed points and the local states of a PFVS of its interaction graph $G(f)$, allowing to conclude that the maximum number of fixed points of a RBN $f$ is $2^{\tau^+(G(f))}$.  A simple exercise shows that this result is also valid in general Boolean networks \cite{richard2019positive}.
In this way, we develop an algorithm that considers all the possible states of a given PFVS and efficiently verifies, by application of $f$, which of them produces a fixed point for the network. This is achieved thanks to the dynamical behavior of a BN without positive cycles described in \Cref{teo:maestra}. Previously, we give some definitions and results.

\begin{definicion}\label{def:fminus}
Let $\bn$ be a BN, $P$ a $\PFVS$ of $G(f)$ and $a: P \to \B$  a function, which can be represented by a vector $a \in \B^{|P|}$. We define the Boolean network  \mbox{$\bn[\funca]$} as follows:
\[
    \forall v \in [n],  \forall x \in \B^n,\   \funca_v(x) = 
    \begin{cases} 
        a(v) & \text{if } v\in P,\\
        f_v(x) & \text{if }v\notin P.
    \end{cases}
\]
\end{definicion}

\begin{ejemplo}\label{exa:fminus}
\Cref{fig:fminus} shows an example of a BN $f$ and a BN $\funca$, where $a(1) = a(3) = 0$ and $a(2) = 1$, according to \Cref{def:fminus}. Dark gray vertices represent $P$.

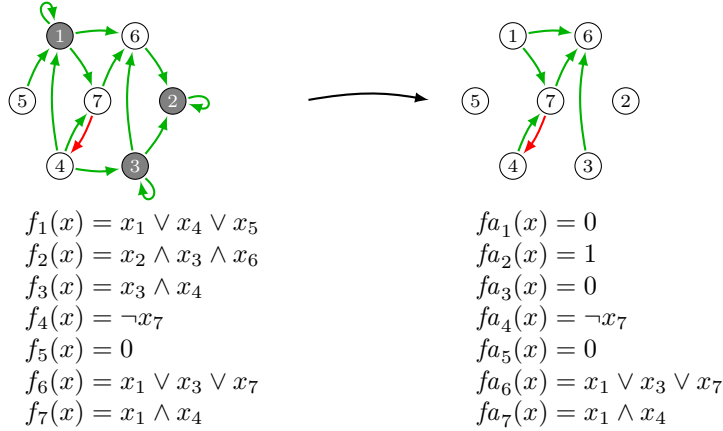
\begin{figure}[!h]
\centering
\begin{tabular}{lcl}
\begin{tikzpicture}[baseline=(7.0)]
\coordinate (c1) at (120:\unitlength);
\coordinate (c2) at (0:\unitlength);
\coordinate (c3) at (300:\unitlength);
\coordinate (c4) at (240:\unitlength);
\coordinate (c5) at (180:\unitlength);
\coordinate (c6) at (60:\unitlength);
\coordinate (c7) at (0,0);
\coordinate (t1) at (0:1.5\unitlength);
\coordinate (t2) at (0:2.5\unitlength);
\foreach \x in {1,2,3}{
	\node[vertex, semi](\x) at (c\x) {\x};
}
\foreach \x in {4,6,5,7}{
	\node[vertex](\x) at (c\x) {\x};
}
\Bucle[verde]{1}{120}{};
\Bucle[verde]{2}{0}{};
\Bucle[verde]{3}{-60}{};
\path[arcos,bend left=10,verde]
	(1) edge (6)
	(1) edge (7)
	(3) edge[bend right=10] (2)
	(3) edge (6)
	(4) edge (1)
	(4) edge[bend right=10] (3)
	(4) edge (7)
	(5) edge (1)
	(6) edge (2)
	(7) edge[rojo] (4)
	(7) edge (6);
\end{tikzpicture} &
\begin{tikzpicture}
\path[arcos]
    (-0.8cm,0) edge[bend left=10] (0.8cm,0);
\end{tikzpicture} &
\begin{tikzpicture}[baseline=(7.0)]
\foreach \x in {1,...,7}{
	\node[vertex](\x) at (c\x) {\x};
}
\path[arcos,bend left=10,verde]
	(1) edge (6)
	(1) edge (7)
	(3) edge (6)
	(4) edge (7)
	(7) edge[rojo] (4)
	(7) edge (6);
;
\end{tikzpicture}\\
\begin{tabular}{l}
$f_1(x) = x_1 \lor x_4 \lor x_5$\\
$f_2(x) = x_2 \land x_3 \land x_6$\\
$f_3(x) = x_3 \land x_4$\\
$f_4(x) = \lnot x_7$\\
$f_5(x) = 0$\\
$f_6(x) = x_1 \lor x_3 \lor x_7$\\
$f_7(x) = x_1 \land x_4$\\
\end{tabular} & &
\begin{tabular}{l}
$\funca_1(x) = 0$\\
$\funca_2(x) = 1 $\\
$\funca_3(x) =0 $\\
$\funca_4(x) = \lnot x_7$\\
$\funca_5(x) = 0$\\
$\funca_6(x) =x_1 \lor x_3 \lor x_7$\\
$\funca_7(x) =  x_1 \land x_4$\\
\end{tabular} \\
\end{tabular}
\caption{A Boolean network $f$ and a \emph{network without positive cycles} $\funca$.}
\label{fig:fminus}
\end{figure}
\end{ejemplo}

Note that $G(\funca)= G(f)- \{(u,v) \in A(G(f)):\ v \in P\}$. Hence, $\funca$ is a BN without positive cycles.
As a direct result, we have the following proposition.
\begin{proposicion}\label{prop:ex-lema}
Let $\bn$ be a BN, $P \neq \emptyset$ a PFVS of $G(f)$ and $a: P \to \B$ a function.  $x \in \B^n$ is a fixed point of $f$ such that $\forall v \in P, x_v = a(v)$ if and only if:
\begin{enumerate}
\item $x$ is a fixed point of $\funca$,
\item $\forall v \in P, f_v(x) = a(v)$.
\end{enumerate}
\end{proposicion}

\begin{demostracion}
$(\Rightarrow)$ Let $x \in \B^n$ be a fixed point of $f$ such that $\forall v \in P, x_v = a(v)$, then, as $x$ is a fixed point of $f$, then $\forall v, f_v(x) = x_v$. Hence, by \Cref{def:fminus}, if $v \notin P, \funca_v(x) = f_v(x) = x_v$, and if $v \in P, \funca_v(x) = a(v) = x_v$. Therefore, $\forall v, \funca_v(x) = x_v$, thus, $x$ is a fixed point of $\funca$ and, by hypothesis, condition (2) also holds. 

$(\Leftarrow)$ If $x \in \B^n$ is a fixed point of $\funca$, then $\forall v, \funca_v(x) = x_v$. Hence, by \Cref{def:fminus}, if $v \notin P, x_v = \funca_v(x) = f_v(x)$.

Moreover, If $\forall v \in P, x_v = \funca_v(x) = a(v) = f_v(x)$.

Therefore, $\forall v, x_v = f_v(x)$, then, $x$ is a fixed point of $f$ and $\forall v \in P, x_v = a(v)$.
\end{demostracion}

In this way, we use \Cref{prop:ex-lema} to define \Cref{algo1}, which finds the fixed points of a given BN.
\begin{proposicion}
Given $\bn$ be a BN and $P$  a PFVS of $G(f)$, \Cref{algo1} finds the set of fixed points of $f$ in time  $O(n^22^{|P|})$.
\end{proposicion}

\begin{demostracion}
The correctness of the algorithm is direct from \Cref{prop:ex-lema} and  \Cref{cor:linear}. In particular, if $P \neq \emptyset$, the instruction $x \gets \funca^{\langle n\rangle}(\vec{0})$ obtains a fixed point candidate (after $n$ executions of $\funca$, for some $a \in \B^{|P|}$) that is checked in the next line according to \Cref{prop:ex-lema}. Otherwise, i.e. $f$ has no positive cycles, $x \gets f^{\langle n\rangle}(\vec{0})$ obtains a fixed point candidate (after $n$ executions of $f$) that is checked in the next line according to \Cref{cor:linear}. Finally, the total number of operations, corresponding mainly to  the applications of the local activation functions, in the worst case is $O(n^22^{|P|})$.
\end{demostracion}

\begin{algoritmo}[algo1]{\Abfp[:]}
\KwIn{$f$ a BN with $n$ components and $P$ a PFVS of $G(f)$.}
\KwOut{$S$ the set of fixed points of $f$.}
$S \gets \emptyset$\;
\uIf{$P \neq \emptyset$}{
\ForEach{$a \in \B^{|P|}$}{
$x \gets \funca^{\langle n\rangle}(\vec{0})$\;
\If{$(\funca(x) = x) \land (\forall u \in P, f_u(x) = a(u))$}{$S \gets S \cup \{x\}$\;}
}}
\Else{
$x \gets f^{\langle n\rangle}(\vec{0})$\;
\lIf{$(f(x) = x)$}{$S \gets \{x\}$}
}
\Return{$S$}
\end{algoritmo}

In order to make the \Cref{algo1} faster, and following the ideas introduced in \cite{aracena2018fixing},  we use a sequential update schedule that allows a faster convergence to the fixed points when they exist. By \Cref{teo:maestra}, the BNs without positive cycles has a sequential update schedule such that they convergence to the only fixed point, when there exists, in only one step. However, determining such a schedule can be as difficult as finding the fixed points of the network, because it depends on the value of each local activation functions as shown in \Cref{exa:no_positive_cycles}. We propose to use a sequential update schedule which depends only on the interaction graph structure of input network. More precisely, given a FVS $F$ and a PFVS $P$ contained in it, we determine in polynomial time a sequential scheme such that the global activation function applied with  this schedule in each iteration fixes at least the value of one vertex in $F\setminus P$ (in the case that this is possible), then the network is updated once more to fix the remaining vertices. Thus, the number of applications in the new algorithm is reduced from $n$ to $|F|-|P|+1$.

\begin{definicion}\label{def:sf}
Given $\bn$ a BN, $F$ a FVS of $G(f)$ and $P \subseteq F$ a PFVS of $G(f)$,  We say that a permutation $\pi = (\pi_1,\pi_2,\dots,\pi_n)$ on the set $[n]$ is an \emph{order compatible with $F$ and $P$} if it satisfies the following properties:
\begin{itemize}
\item $\forall \pi_i \in (F-P), \forall \pi_j \notin (F-P),\ i > j$.
\item $\forall \pi_i \in P, \forall \pi_j \notin P,\ i < j$.
\item $\forall \pi_i, \pi_j \notin F,\ ( \pi_i, \pi_j) \in A(G(f)) \Longrightarrow i < j$.
\end{itemize}
\end{definicion}

\begin{ejemplo}
\Cref{fig:sf} shows examples of orders compatible with $F$ and $P$. Dark gray vertices represent the PFVS $P$ and light gray vertices represent $F \setminus P$.

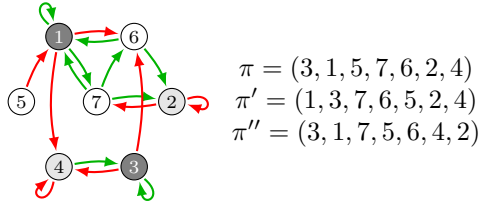
\begin{figure}[htbp]
\centering
\begin{tikzpicture}
\node[vertex, semi] (1) at (c1) {1};
\node[vertex, light] (4) at (c4) {4};
\node[vertex, semi] (3) at (c3) {3};
\node[vertex] (6) at (c6) {6};
\node[vertex] (5) at (c5) {5};
\node[vertex] (7) at (c7) {7};
\node[vertex, light] (2) at (c2) {2};
\Bucle[verde]{1}{120}{};
\Bucle[rojo]{2}{0}{};
\Bucle[verde]{3}{300}{};
\Bucle[rojo]{4}{240}{};
\path[arcos,verde,bend left=10]
    (6) edge (1)
    (6) edge (2)
    (7) edge (6)
    (7) edge (1)
    (7) edge (2)
    (4) edge (3)
    (1) edge (7)
;
\path[arcos,rojo,bend left=10]
    (1) edge[bend right=10] (4)
    (3) edge (4)
    (3) edge[bend right=10] (6)
    (5) edge (1)
    (2) edge (7)
    (1) edge (6)
;
\node[right] (l) at (1.5\unitlength,0) {
$\begin{array}{c}
\pi = (3,1,5,7,6,2,4)\\
\pi' = (1,3,7,6,5,2,4)\\
\pi'' = (3,1,7,5,6,4,2)\\
\end{array}$
};
\end{tikzpicture}

\caption{A digraph $G(f)$ and orders compatible with $F$ and $P$. Set $P$ is denoted for dark gray vertices and set $F$ for light gray vertices.}
\label{fig:sf}
\end{figure}
\end{ejemplo}

\begin{proposicion}\label{prop:maestra-dos}
Let $\bn$ be a BN, such that the interaction graph $G(f)$ does not have positive cycles, $F$ be a FVS of $G(f)$, $\pi$ be an order compatible with $F$ and $y \in \B^n$. $y$ is the unique fixed point of $f$ if and only if \mbox{$\forall x \in \B^n, (f^{\pi})^{\langle|F|+1\rangle}(x) = y$}.
\end{proposicion}
\begin{demostracion}
$(\Rightarrow)$ Let us suppose that $y \in \B^n$ is the unique fixed point of $f$. Without loss of generality, let us suppose that $\pi = (1,2,\dots,n)$. Then, by \Cref{teo:maestra}, there exist \mbox{$I_1(f) \subsetneq I_2(f) \subsetneq \dots \subsetneq I_{k-1}(f) \subsetneq I_k(f) = [n]$} such that:
\[ \forall j \in \Set{1,\dots,k}, \forall v \in I_j(f), \forall t \geq j, \forall x \in \B^n, f_v^{\langle t\rangle}(x) = y_v.\]
Now, we prove that by iterating $f$ according to the order $\pi$, the $n$ vertices fix their value in $m$ iterations (with $m \leq |F|+1$).

To perform this, let us consider the following sequence of indices $o_i$ defined as follows:
\begin{align*}
o_1 &= \min\Set{j \in \Set{1,\dots,k} : (I_j(f) \cap F) \neq \emptyset}, \\
\forall p \geq 2, o_p &= \min\Set{j > o_{p-1} : ((I_j(f) \setminus I_{j-1}(f)) \cap F \neq \emptyset) \lor (j = k)}.
\end{align*}

Notice that $o_1 < o_2 < \dots < o_m = k$ and $m \leq |F|+1$.

We prove that $\forall i \in \Set{1,\dots,m}, \forall u \in I_{o_i}, \forall t \geq i, \forall x \in \B^n, (f^\pi)_u^{\langle t\rangle}(x) = y_u$.

By contradiction, let us suppose that:
\[ \exists i \in \Set{1,\dots,m}, \exists u \in I_{o_i}, \exists t \geq i, \exists x \in \B^n, (f^\pi)_u^{\langle t\rangle}(x) \neq y_u.\]
Let:
\begin{align*}
i_* &= \min\Set{ i \in \Set{1,\dots,m} : \exists u \in I_{o_i},  \exists t \geq i, \exists x \in \B^n, \,  (f^\pi)_u^{\langle t\rangle}(x) \neq y_u} \text{ and}\\
l_* &= \min\Set{ l \leq o_{i_*} : \exists u \in I_l, \exists t \geq i_*, \exists x \in \B^n, \,  (f^\pi)_u^{\langle t\rangle}(x) \neq y_u}.
\end{align*}
Let $u_* \in I_{l_*} \setminus I_{l_*-1}$ be such that $\exists t \geq i_*, \exists x \in \B^n, (f^\pi)^{\langle t\rangle}_{u_*}(x) \neq y_{u_*}$. Then, by \Cref{def:setI}, $\forall t \geq i_*, \forall x \in \B^n$:
\begin{align*}
(f^\pi)^{\langle t\rangle}_{u_*}(x) &= f^\pi_{u_*}((f^\pi)^{\langle t-1\rangle}(x))\\
&= f_{u_*}((f^\pi)^{\langle t\rangle}_i(x): i < u_*; (f^\pi)^{\langle t-1\rangle}_i : i \geq u_*).
\intertext{Notice that $l_* > 1$, then:}
(f^\pi)^{\langle t\rangle}_{u_*}(x) &= f_{u_*}((f^\pi)^{\langle t\rangle}_v(x) : v < u_* \land v \in I_{l_*-1} ; (f^\pi)^{\langle t\rangle}_v(x) : v < u_* \land v \notin I_{l_*-1} ;\\
&\qquad (f^\pi)^{\langle t-1\rangle}_v(x) : v \geq u_* \land v \in I_{l_*-1} ; (f^\pi)^{\langle t-1\rangle}_v(x) : v \geq u_* \land v \notin I_{l_*-1}).
\intertext{By definition of $l_*$, if $v \in I_{l_*-1}$, then $(f^\pi)^{\langle t\rangle}_v(x) = y_v$. Moreover, If $v \geq u_*$, then $v \in F$ and, therefore, if $v \in F$ and $v \in I_{l_*-1}$, then $v \in I_{o_{(i_*-1)}}$ and, thus, if $v\geq u_*$ and $v \in I_{l_*-1}$, then $(f^\pi)^{\langle t-1\rangle}_v(x) = y_v$. Also, if $i_* = 1$, then $\forall l \leq o_1, F \cap I_{l-1} = \emptyset$:}
(f^\pi)^{\langle t\rangle}_{u_*}(x) &= f_{u_*}(y_v : v < u_* \land v \in I_{l_*-1} ; (f^\pi)^{\langle t\rangle}_v(x) : v < u_* \land v \notin I_{l_*-1} ;\\
&\qquad y_v : v \geq u_* \land v \in I_{l_*-1} ; (f^\pi)^{\langle t-1\rangle}_v : v \geq u_* \land v \notin I_{l_*-1})\\
&= f_{u_*}(y_v : v \in I_{l_*-1} ; (f^\pi)^{\langle t\rangle}_v(x) : v < u_* \land v \notin I_{l_*-1} ;\\
&\qquad (f^\pi)^{\langle t-1\rangle}_v(x) : v \geq u_* \land v \notin I_{l_*-1}).
\end{align*}
By \Cref{def:setI}, $\forall t \geq i_*, \forall x \in \B^n, (f^\pi)^{\langle t\rangle}_{u_*}(x) = y_{u_*}$, which is a contradiction.

Therefore, in at most $m$ iterations of $f^\pi$, all vertices in $I_{o_m}(f) = I_k(f)$ are fixed.

$(\Leftarrow)$ Let us suppose that $\forall x \in \B^n, (f^\pi)^{\langle|F|+1\rangle}(x) = y$, then:
\[ f^\pi(y) = f^\pi((f^\pi)^{\langle|F|+1\rangle}(x)) = (f^\pi)^{\langle|F|+2\rangle}(x) = (f^\pi)^{\langle|F|+1\rangle}(f^\pi(x)) = y. \]
Since $f^\pi(y) = y$, then, $y$ is a fixed point of $f$.
\end{demostracion}

\begin{teorema}
Let $F$ be a FVS of $G(f)$ and $P \subseteq F$ a PFVS of $G(f)$, \Cref{algo0} finds the set of fixed points of $f$ in time $O((|F|-|P|+1) n 2^{|P|})$.
\end{teorema}
\begin{demostracion}
As the \Cref{algo1}, the correctness of this algorithm is based on the fact that its execution demonstrates each of the necessary conditions according to the \Cref{prop:ex-lema}. The difference is that the way to find the fixed point of $\funca$ (for some $a \in \B^{|P|}$) is done according to \Cref{prop:maestra-dos}, i.e. instead of executing $n$ times $\funca$, $\funca^{\pi}$ is executed $|F|-|P|+1$ times. For this reason, the complexity of this algorithm is $O((|F|-|P|+1) n 2^{|P|})$.
\end{demostracion}
\begin{algoritmo}[algo0]{\Afp[:]}
\KwIn{$f$ a BN with $n$ components, $F$ a FVS of $G(f)$, $P \subseteq F$ a PFVS of $G(f)$, $\pi$ an order compatible with $F$ and $P$.}
\KwOut{$S$ the set of fixed points of $f$.}
$m \gets |F - P|$\;
$S \gets \emptyset$\;
\uIf{$P \neq \emptyset$}{
\ForEach{$a \in \B^{|P|}$}{
$x \gets (\funca^{\pi})^{\langle m+1\rangle}(\vec{0})$\;
\If{$(\funca(x) = x) \land (\forall u \in P, f_u(x) = a(u))$}{$S \gets S \cup \{x\}$\;}
}}
\Else{
$x \gets (f^{\pi})^{\langle m+1\rangle}(\vec{0})$\;
\lIf{$(f(x) = x)$}{$S \gets S \cup \{x\}$}
}
\Return{$S$}
\end{algoritmo}

\subsection{Building a PFVS}
\Afp requires as input a FVS $F$ and a PFVS $P\subseteq F$  of the interaction graph of a  Boolean network. It is known that the problems of finding a minimum FVS and a minimum PFVS in a signed digraph are both NP-complete \cite{Karp,montalva2008complexity}. In this section we propose an polynomial algorithm that allows to find a PFVS (not necessarily minimal) and a minimal FVS containing it. 

Let $G(f)$ be the signed interaction graph of a BN $f$ with $n$ components. Given $(v_1,v_2,\dots,v_n)$ an order over $V(G(f))$, \Cref{algo2} classifies the vertices of $G(f)$ in the following sets:
\begin{itemize}
\item $P$: A set of vertices that is a $\PFVS$ of $G(f)$.
\item $O$: A set of vertices such that $P\cup O$ is a minimal $\FVS$ of $G(f)$.
\item $R$: The rest of the vertices of $G(f)$.
\end{itemize}
In addition, the algorithm considers the following auxiliar sets:
\begin{itemize}
\item $Y$: A set of vertices that covers some circuit.

\item $U$: The vertices that have not yet been assigned to any set.
\end{itemize}

The operation of the algorithm is as follows:

First, we classify vertices with positive loop directly into $P$. Subsequently, the negative loops are removed from the arcs of $G(f)$. 

Then, the first phase begins. All vertices that are not in $P$ or $R$, are incorporated in $U$ to be visited. Then, the first vertex in $U$ (according to the input order) is incorporated into $R$ (i.e., it is discarded from the PFVS) and if this vertex originally had a negative loop, it is incorporated into $O$ (since that vertex has to be part of the FVS). Subsequently, for each vertex $v \in  U \cup Y$, $G''$ (the subdigraph induced by $R \cup \Set{v}$) is calculated, where, if $G''$ has a circuit, $v$ is removed from $U$ and from $Y$ (since $v$ covers some circuit in $G(f)$), and then, if $G''$ has a positive circuit, $v$ is incorporated in $P$ (since $v$ is the only unclassified vertex of the positive circuit of $G''$), and otherwise, $v$ is incorporated into $Y$ (since it covers some circuit, but no positive circuit is known, to add $v$ to $P$). Once $U$ is empty, the first phase is terminated.

If at the end of the first phase, there are vertices that are neither in $P$ nor in $R$ (for example, they are in $Y$), a second phase begins. The only difference of this phase (and subsequent ones) with respect to the first phase, is that if a vertex is sent to $R$ it is automatically sent to $O$ (since this phase, all unclassified vertices cover some circuit, but being discarded, it means that they do not cover any positive circuit, so they are part of the FVS).

If at the end of the second phase some unclassified vertices remain, a third phase begins under the same conditions of the second phase, and so on.

Notice that, the constructed PFVS is not minimal, since a vertex can be included in $P$ because it covers a positive circuit (not necessarily a positive cycle), but the FVS is minimal, because the algorithm works looking for cycles in $G(f)-P-O$, in this way, once we add a vertex to $P\cup O$ every cycle covered for this vertex it is eliminated of $G(f)-P-O$, so the only reason because a vertex is included in $P\cup O$ is that there exists a cycle that is not covered for another vertex in $P\cup O$, therefore $P\cup O$ is a minimal FVS.

\begin{algoritmo}[algo2]{\Apfvs[:]}
\KwIn{The signed interaction graph $G(f)$ of a BN $f$ with $n$ components and
$(v_1,v_2,\dots,v_n)$ an ordered sequence of $V(G(f))$.}
\KwOut{A PFVS $P$ of $G(f)$, and a FVS $F$ of $G(f)$ such that $P \subseteq F$.}

\tcp{Initialization}
$V^{\oplus} \gets \{u \in V(G(f)): (u,u) \text{ is a positive arc of } G(f) \}$\;
$V^{\ominus} \gets \{u \in V(G(f)): (u,u) \text{ is a negative arc of } G(f)\}$\;
$G' \gets G(f) - \{(u,u) \in A(G(f)): u \in V^{\ominus}\}$\;
$P \gets V^{\oplus}$; $O \gets \emptyset$; $R \gets \emptyset$\;
$\mathrm{phase} \gets 1$\;
\While{$(P \cup R) \neq V(G(f))$}{
$U \gets V(G(f)) \setminus (P \cup R)$\;
$Y \gets \emptyset$\;
\While{$U \neq \emptyset$}{
$u \gets $ vertex in $U$ with lowest index\;
$U \gets U \setminus \{u\}$; $R \gets R \cup \{u\}$\;
\lIf{($\mathrm{phase} > 1) \lor (u \in V^{\ominus})$}{$O \gets O \cup \{u\}$}
\tcp{Force-subroutine}
\ForEach{$v \in U \cup Y$}{
$G'' \gets G'[R \cup \{v\}]$\;
\If{$G'' \text{ has a circuit}$}{
$U \gets U \setminus \{v\}$; $Y \gets Y \setminus \{v\}$\;
\lIf{$G'' \text{ has a positive circuit}$}{$P \gets P \cup \{v\}$}
\lElse{$Y \gets Y \cup \{v\}$}
}}
\tcp{end Force-subroutine}
}
$\mathrm{phase} \gets \mathrm{phase} + 1$\;}
$F \gets P \cup O$\;
\Return{$P$ and $F$}
\end{algoritmo}

The following algorithm is the way we implement the force-subroutine.  A vertex $v \in V(G(f))$ is an \emph{ancestor} of a vertex $u$ if $v \in R$ and exists a path $P_{vu}$ from $v$ to $u$ such that $\forall w \in P_{vu}, w \in R$. The set of ancestors of $u$ is denoted by $\anc(u)$. A vertex $v \in V(G(f))$ is a \emph{descendant} of a vertex $u$ if $v \in U \cup Y$ and exists a path $P_{vu}$ such that $\forall w \in V(P_{vu}), w \neq u$ implies $w \in R$. The set of descendants of $u$ is denoted by $\dec(u)$. The sets $\bef(u)$ and $aft(u)$ contain the elements related to $u$ (the last vertex added to $R$) that can form a circuit. If there is an arc that goes from a vertex $a \in \aft(u)$ to a vertex $b \in \bef(u)$, it means that there is a circuit formed by:
\begin{itemize}
\item The path from $b$ to $u$,
\item The path from $u$ to $a$ and
\item The arc from $a$ to $b$.
\end{itemize}
The sign of this circuit will determine how the vertex $a$ is classified (If the circuit is positive, $a$ is added to $P$. If the circuit is negative, $a$ is added to $Y$).\\
The sets $\anc(u), \dec(u), \bef(u)$ and $\aft(u)$ contain elements of the form $(v,\sigma)$, where $v$ is a vertex, and $\sigma$ is a sign ($+1$ or $-1$). Let $x$ be an element of any of these sets, we will denote $v(x)$ to the vertex associated with the element $x$, and we will denote $\sigma(x)$ to the sign associated with the element $x$.

\begin{algoritmo}[algo3]{Force-subroutine}
\tcp{During initialization}
\ForEach{$v \in V(G(f))$}{
$\anc(v) \gets \emptyset$; $\dec(v) \gets \emptyset$\;
}
\tcp{Force-subroutine, ``u'' is the selected vertex in PFVS-Algorithm}
$\bef(u) \gets \anc(u) \cup \Set{(u,+1)}$\;
$\aft(u) \gets \emptyset$\;
\ForEach{$v \in N^+(u)$}{
\lIf{$v \in U$}{$\aft(u) \gets \aft(u) \cup \Set{(v,\sigma(u,v))}$}
\lElse{$\aft(u) \gets \aft(u) \cup \dec(v)$}
}
\ForEach{$(a,b) \in \aft(u) \times \bef(u)$}{
\eIf{$(v(a),v(b)) \in A(G)$}{
$U \gets U \setminus \Set{a}$\;
$\sigma \gets \sigma(a) \cdot \sigma(b) \cdot \sigma(v(a),v(b))$\;
\lIf{$\sigma = +$}{$P \gets P \cup \Set{d}$}
\Else{$Y \gets Y \cup \Set{d}$\;
$\dec(b) \gets \dec(b) \cup \Set{(a,\sigma_d \cdot \sigma_a)}$\;
$\anc(a) \gets \anc(a) \cup \Set{(b,\sigma_d \cdot \sigma_a)}$\;}
}{
$\dec(b) \gets \dec(b) \cup \Set{(a,\sigma_d \cdot \sigma_a)}$\;
$\anc(a) \gets \anc(a) \cup \Set{(b,\sigma_d \cdot \sigma_a)}$\;
}
}
\end{algoritmo}

\subsubsection{Heuristics for the order of the vertices}

Different orders of the vertices as input of \Apfvs can generate different PFVS as outputs, so it may be necessary to define some heuristics to choose an appropriate order of the vertices. In this paper we use random order and Min-order defined as follows: 

\begin{itemize}
\item Min-order:  the vertices are ordered according:
\begin{enumerate}
\item Their degrees, from lowest to highest.
\item If two or more vertices have equal degree then by in-degree, from lowest to highest.	
\end{enumerate}
\end{itemize}

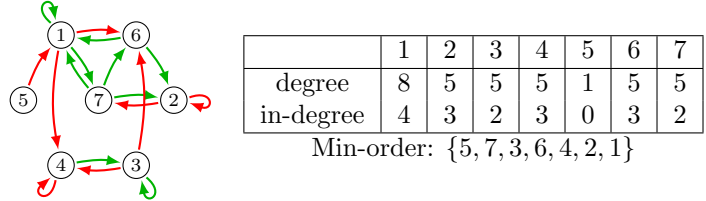
\begin{figure}[htbp]
\centering
\begin{tikzpicture}
\node[vertex] (1) at (c1) {1};
\node[vertex] (4) at (c4) {4};
\node[vertex] (3) at (c3) {3};
\node[vertex] (6) at (c6) {6};
\node[vertex] (5) at (c5) {5};
\node[vertex] (7) at (c7) {7};
\node[vertex] (2) at (c2) {2};
\Bucle[verde]{1}{120}{};
\Bucle[rojo]{2}{0}{};
\Bucle[verde]{3}{300}{};
\Bucle[rojo]{4}{240}{};
\path[arcos,verde,bend left=10]
    (6) edge (1)
    (6) edge (2)
    (7) edge (6)
    (7) edge (1)
    (7) edge (2)
    (4) edge (3)
    (1) edge (7)
;
\path[arcos,rojo,bend left=10]
    (1) edge[bend right=10] (4)
    (3) edge (4)
    (3) edge[bend right=10] (6)
    (5) edge (1)
    (2) edge (7)
    (1) edge (6)
;

\node[right] (l) at (1.8\unitlength,0){
\begin{tabular}{|c|c|c|c|c|c|c|c|}
\hline
& 1 & 2 & 3 & 4 & 5 & 6 & 7\\
\hline
degree & 8 & 5 & 5 & 5 & 1 & 5 & 5\\
in-degree & 4 & 3 & 2 & 3 & 0 & 3 & 2\\
\hline
\multicolumn{8}{c}{Min-order: $\Set{5,7,3,6,4,2,1}$}\\
\end{tabular}};
\end{tikzpicture}
\caption{A signed interaction graph $G(f)$ and the order of its vertices according to the Min-order.}
\label{fig:sort}
\end{figure}

\Cref{fig:sort} shows an example Min-order.

\section{Results}
\begin{figure}[hp!]
\begin{center}
\begin{tikzpicture}
\node (f6) at (0,0) {\includegraphics[width=\linewidth]{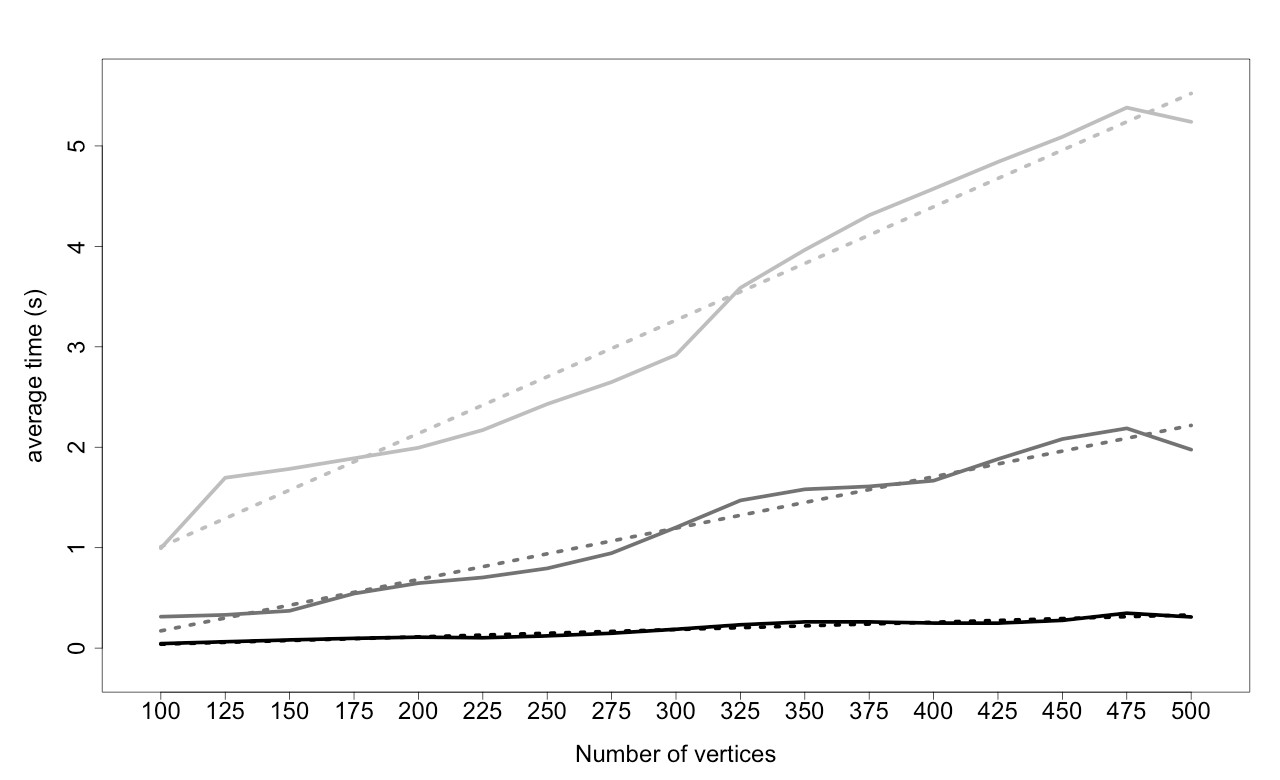} };
\node[below] (t6) at (f6.north) {\bf\small for different  network sizes $\boldsymbol{(\tau=15)}$};
\node[above,inner sep=0pt,outer sep=0pt] (t6a) at (t6.north) {\bf\small Time performance of \Afp };
\node (l15) at (0.4\linewidth, 3cm) {\small $\tau^+=15$};
\node (l10) at (0.4\linewidth, 0cm) {\small $\tau^+=10$};
\node (l5) at (0.4\linewidth, -2.2cm) {\small $\tau^+=5$};
\end{tikzpicture}
\end{center} 
\caption{Results of \Afp.}\label{fig:FPAlgoA}
\end{figure}

\begin{figure}[hp!]
\begin{center}
\begin{tikzpicture}
\node (f7) at (0,0) {\includegraphics[width=\linewidth]{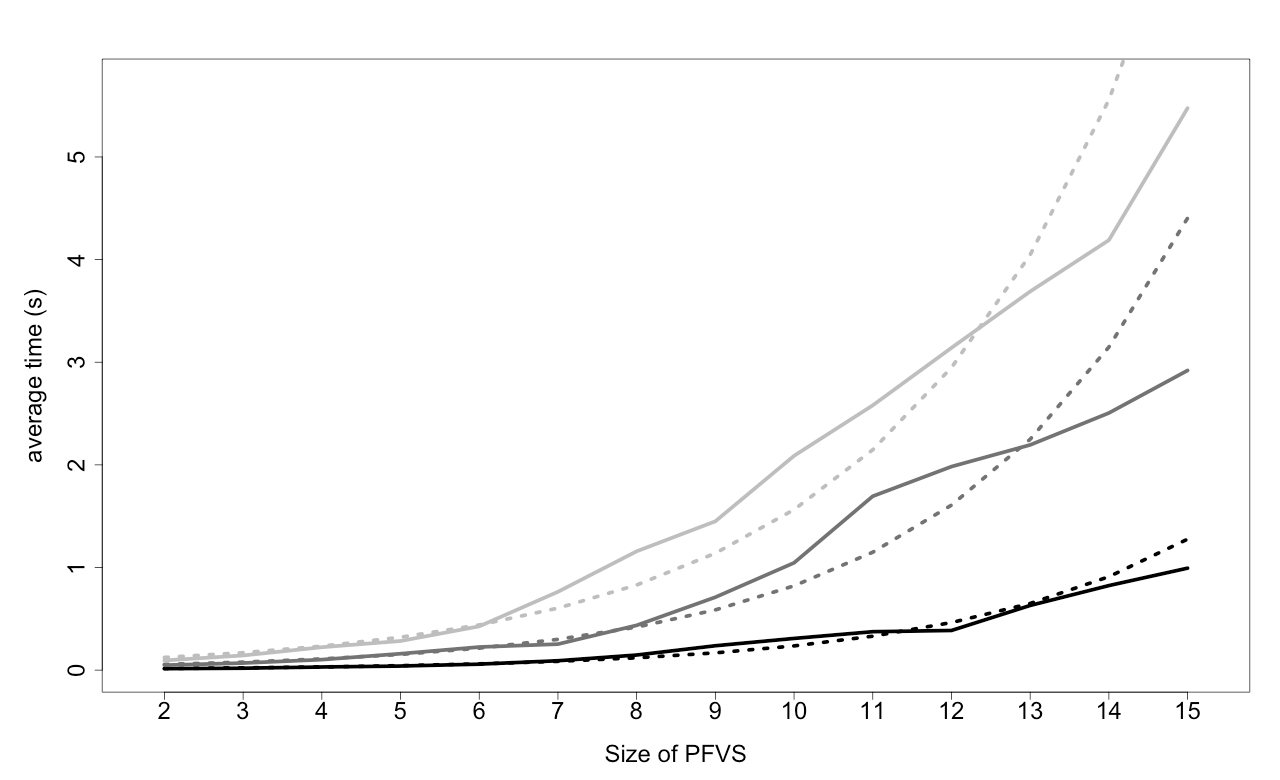} };
\node[below] (t7) at (f7.north) {\bf\small for different  PFVS sizes $\boldsymbol{(\tau=15)}$};
\node[above,inner sep=0pt,outer sep=0pt] (t7a) at (t7.north) {\bf\small Time performance of \Afp };
\node (l5) at (0.4\linewidth, 2cm) {\small $n=500$};
\node (l3) at (0.4\linewidth, -0.3cm) {\small $n=300$};
\node (l1) at (0.4\linewidth, -1.5cm) {\small $n=100$};
\end{tikzpicture}
\end{center}
\caption{Results of \Afp.}\label{fig:FPAlgoB}
\end{figure}

\begin{figure}[ht!]
\begin{center}
\begin{tikzpicture}
\node (f8) at (0,0) {\includegraphics[width=\linewidth]{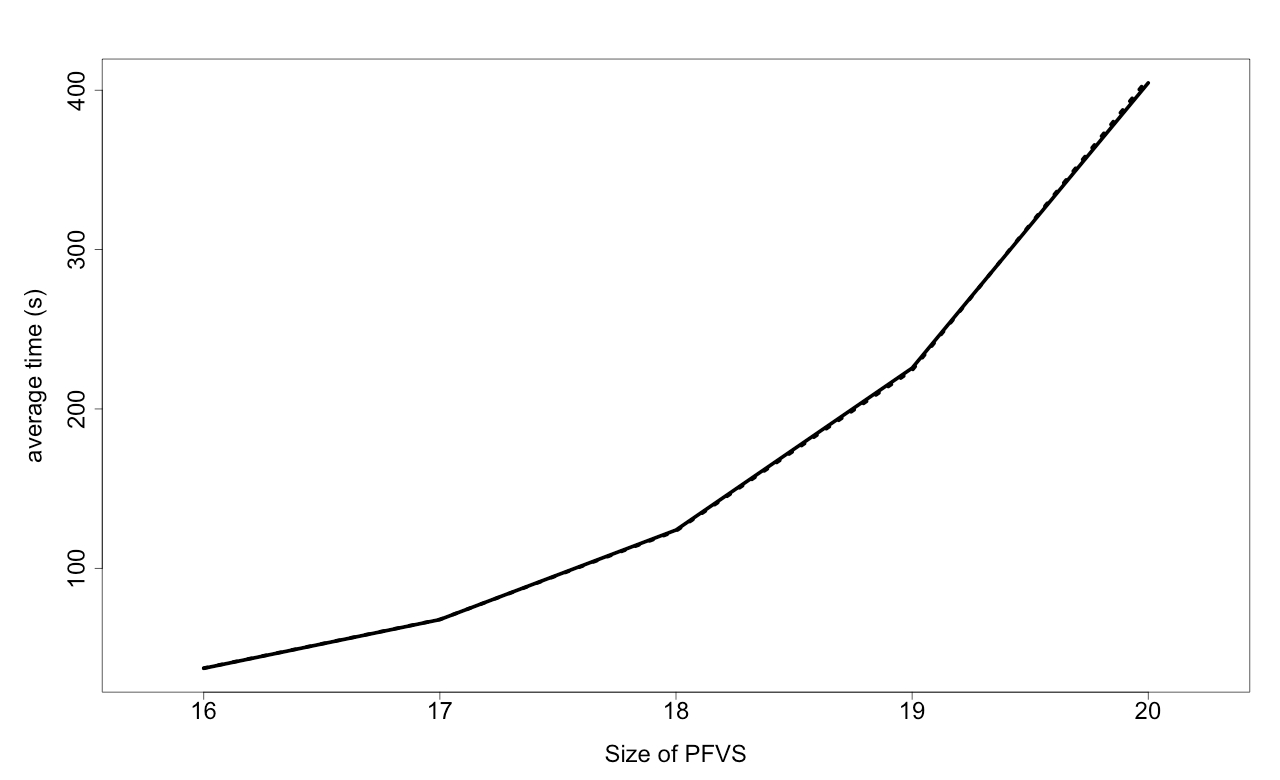} };
\node[below] (t8) at (f8.north) {\bf\small for different PFVS sizes $\boldsymbol{(n=300,\tau=30)}$};
\node[above,inner sep=0pt,outer sep=0pt] (t8a) at (t8.north) {\bf\small Time performance of \Afp };
\end{tikzpicture}
\end{center}
\caption{Results of \Afp.}\label{fig:FPAlgoC}
\end{figure}

We first tested the running time of \Afp in random BNs with a fixed transversal number $\tau$ and with different numbers of components $n$ and positive transversal numbers $\tau^+$. 
 The tested networks were constructed in such a way that a minimum FVS and a minimum PFVS were both known a priori, i.e. we did not use \Apfvs in these cases. The average times obtained in milliseconds with one hundred networks tested in each case, are shown in \Crefrange{fig:FPAlgoA}{fig:FPAlgoC}. We can see that the time performance of \Afp grows exponentially with the value of $\tau^+$ and polynomially with the size of the network.  


\begin{figure}[ht!]
\begin{center}
\begin{tikzpicture}
\node (f9) at (0,0) {\includegraphics[width=\linewidth]{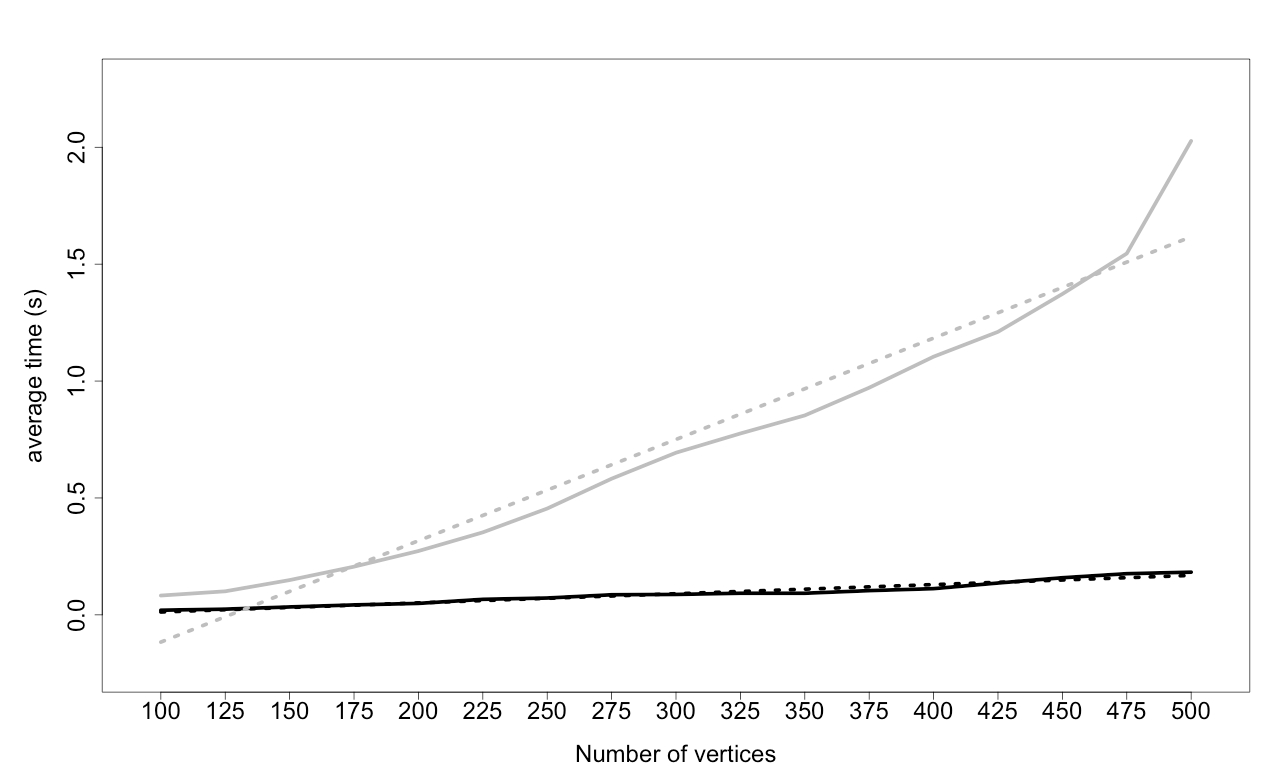} };
\node[below] (t9) at (f9.north) {\bf\small for different  network sizes $\boldsymbol{(\tau=15,\tau^+=5)}$};
\node[above,inner sep=0pt,outer sep=0pt] (t9a) at (t9.north) {\bf\small Time performance of \Apfvs };
\node (min) at (0.4\linewidth, -2cm) {\small min};
\node (rand) at (0.4\linewidth, 1.5cm) {\small rand};
\end{tikzpicture}
\end{center}
\caption{Results of \Apfvs.}\label{fig:PFVSAlgoA}
\end{figure}

The results of \Apfvs using different heuristic to chose the order of the vertices are presented in \Cref{fig:PFVSAlgoA}. We can see that running time of the algorithm depends mainly on the size of the network and not on the sizes of the positive feedback vertex sets. Besides, in random networks as well as in networks from the literature described in \Cref{table:results} the performance of the min-order heuristic is better than the random one.  

On the other hand, the correctness and time performance of \Afp and \Apfvs were also tested with networks from the literature where the sets of fixed points obtained coincides with those published. The results of these tests, shown in \Cref{table:results}, were obtained with a PC 3.60GHz Intel Core i7 processor with 16GB of RAM and can be directly checked  with the implementation FixedPointsCollector located at \href{http://www.inf.udec.cl/~lilian/FPCollector/}{\url{www.inf.udec.cl/~lilian/FPCollector/}}. We can observe that the running times obtained in these networks are short, probably due to the small size of their positive feedback vertex sets.   The latter is also observed in another family of published networks, which are shown in \Cref{tab:2}, where the times obtained (using the same PC) are shorter than those obtained by the Veliz-Cuba algorithm in \cite{veliz2014steady} for networks with a small $\tau^+$, however the same does not occur when $\tau^+$ is large.
\begin{table}[ht!]
\caption{Performance of \Afp in real networks.}\label{table:results}
\begin{center}
\newcommand{\fg}{\rowcolor[gray]{0.8}}
\begin{tabular}{lrrrrr}
\hline\fg
\bf Network & $\boldsymbol{n}$ & \bf FP & $\boldsymbol{|P|}$ & \bf T(min) & \bf T(rand) \\ \hline
Cancer cell \cite{yang2013robustness} & 8 & 5 & 4 & 6.952 & 8.248 \\ \fg
Cancer backbone \cite{yang2013robustness} & 8 & 12 & 5 & 7.019 & 7.321 \\
Budding yeast \cite{li2004yeast} & 11 & 7 & 6 & 5.963 & 7.036 \\ \fg
Fission yeast \cite{davidich2008boolean} & 10 & 12 & 4 & 8.980 & 9.416 \\
Arabidopsis Thaliana \cite{sanchez2010arabidopsis} & 13 & 10 & 7 & 8.938 & 8.919 \\ \fg
T-helper cell \cite{mendoza2006method} & 23 & 3 & 3 & 7.407 & 13.512 \\ 
T-cell receptor \cite{klamt2006methodology} & 40 & 1 & 1 & 12.782 & 29.895 \\ \fg
HGF \cite{singh2012boolean} & 66 & 2 & 3 & 17.286 & 61.969 \\
Apostosis model 1 \cite{kobayashi2019design} & 12 & 2 & 2 & 6.923 & 8.181 \\ \fg
Apostosis model 2 \cite{kobayashi2019design} & 12 & 4 & 2 & 6.870 & 9.170 \\
Apostosis model 3 \cite{kobayashi2019design} & 3 & 4 & 2 & 5.143 & 6.174 \\ \fg
Drosophila melanogaster \cite{albert2003topology} & 60 & 10 & 9 & 19.140 & 64.565 \\
Ventral spinal cord \cite{lovrics2014boolean} & 8 & 5 & 4 & 6.751 & 7.126 \\ \hline
\end{tabular}
\end{center}

{\footnotesize $|P|$ is the size the PFVS obtained with min-order.\\
T(min) and T(rand) correspond to the time in milliseconds of the PFVS-Alg.+FP-Alg. using Min-order and random-order (in this case, its considered the execution of $n/2$ random permutations to choose the smallest PFVS), respectively.}
\end{table}

\begin{table}[htp!]
\caption{Time Performance Comparison}\label{tab:2}
\begin{center}
\newcommand{\fg}{\rowcolor[gray]{0.8}}
\begin{tabular}{lrrrrrr}
\hline\fg
& & & \multicolumn{2}{c}{\bf Veliz-Cuba} & \multicolumn{2}{c}{\bf Our algorithm} \\ \fg
\bf Network & $\boldsymbol{n}$ & $|P|$ & \bf mean & \bf stdev. & \bf mean & \bf stdev. \\ \hline
HGF Signaling in Keratinocytes \cite{singh2012boolean} & 68 & 3 & 2.347 & 0.299 & 0.451 & 0.079 \\ \fg
T Cell Receptor Signaling \cite{saez2007logical} & 101 & 1 & 2.355 & 0.337 & 0.347 & 0.054 \\
Signaling in Macrophage Activation \cite{raza2008logic} & 321 & 3 & 2.406 & 0.298 & 0.464 & 0.044 \\ \fg
Yeast Apostosis \cite{kazemzadeh2012boolean} & 73 & 3 & 2.370 & 0.429 & 0.378 & 0.042 \\
Influenza A Virus Replication Cycle \cite{madrahimov2013dynamics} & 131 & 12 & 3.599 & 0.897 & 0.407 & 0.044 \\ \fg
T-LGL Survival network (v. 2011) \cite{saadatpour2011dynamical} & 60 & 10 & 3.152 & 0.786 & 0.580 & 0.041 \\
T-LGL Survival network (v. 2008) \cite{zhang2008network} & 61 & 10 & 1.790 & 1.181 & 0.409 & 0.044 \\ \fg
EGFR \& ErbB Signaling \cite{samaga2009logic} & 104 & 2 & 1.185 & 0.910 & 0.356 & 0.041 \\ 
Signal Transduction in Fibroblasts \cite{helikar2008emergent} & 139 & 39 & 23,19* & 98.42 & DNF & DNF \\ \fg
ErbB (1-4) Receptor Signaling \cite{helikar2013comprehensive} & 247 & 42 & 4186* & 12284 & DNF & DNF\\ \hline
\end{tabular}
\end{center}
{\footnotesize $|P|$ is the size the PFVS obtained with min-order. Times presented in seconds.\\
*Times obtained from \cite{veliz2014steady}. DNF: Tests that did not finish within 3 days}
\end{table}

\newpage
\section{Discussion}
In the modeling of biological systems by Boolean networks usually the interaction graph of the network is known as well as the type of interaction between their components (activation, inhibition). In this way, it seems natural to use this information to determine the fixed points of a network. In this paper, we have constructed \Afp to find the set of fixed points of a Boolean network based mainly on the structure of the positive cycles of its interaction graph and to a lesser extent on the size of the network. This can be considered an improvement to the results obtained by Akutsu et al. in \cite{akutsu1998system}, since the set of states to test is smaller and, in the case that $\tau=\tau^+$, our algorithm works the same way. 
The theoretical foundation of the algorithm is given by \Cref{teo:maestra}, which provides a nice characterization of the dynamical behavior of Boolean networks without positive cycles and with a fixed point.

The efficiency of \Afp depends mainly on the size of the input PFVS. Due to this, it is important to have a PFVS as close to the minimum as possible. In this sense, it is clear that \Apfvs can be improved. One improvement could be done implementing an efficient algorithm to decide the existence of a positive cycle in a signed digraph.  This latter problem, which is equivalent to the existence of an even cycle in a digraph \cite{montalva2008complexity}, is a surprisingly difficult problem and whose algorithmic complexity was unknown for a long time. Although Robertson, Seymour and Thomas finally proved that this decision problem can be solved in polynomial time \cite{robertson1999permanents}, the implementation of such algorithm is not easy to do which makes it one of our future challenges. Another way to improve \Apfvs is the possibility of building a fixed-parameter tractable algorithm to determine a PFVS of minimum size, i.e. an algorithm whose complexity depends mainly on this size and not on the size of the interaction graph. It is known that this type of algorithm exists to solve the minimum FVS problem in digraphs \cite{chen2008fixed}. However, it is an open problem, at our knowledge, in the case of minimum PFVS in signed digraphs. 

In future work it would be important to explore methods to reduce the size of a network that conserve $\tau^+$. Also, since the efficiency of the \Afp depends on the size of a input PFVS, it is important to study the relationship between $\tau^+$ and other parameters of the interaction graph of a Boolean network such as: minimum  and maximum in-degree,  out-degree and  degree distributions.

\section*{Acknowledgements}
We would like to thank Adrien Richard for his review and suggestions for improving the manuscript.
\section*{Funding}
J. Aracena was partially supported by CONICYT-Chile through the  
project AFB170001 of the PIA Program: \emph{Concurso Apoyo a Centros Cient\'ificos y Tecnol\'ogicos de Excelencia con  Financiamiento Basal}. L. Cabrera-Crot is funded by CONICYT-PCHA/Doctorado Nacional/2016-21160885.

\appendix
\section{Example of the algorithms}

\begin{ejemplo}
\Cref{figA1,figA2} show the operation of the FixedPoint-Algorithm with a Boolean network with $\tau^+ = 1$ and $\tau = 3$. Note that in the tables in steps 3 and 5, the purple cells represent the components that are fixed according to \Cref{def:setI}
\end{ejemplo}

\newcommand{\arcosPos}{%
 (2) edge (3)
 (5) edge (6)
 (5) edge (8)
 (4) edge (3)
 (3) edge (5)
 (6) edge (3)
 (7) edge[bend left=10] (1)
 (8) edge[bend left=40] (7)
 (2) edge[bend left=20] (4)
 }
\newcommand{\arcosNeg}{%
 (3) edge (1)
 (8) edge (4)
 (1) edge[bend left=10] (7)
 (7) edge[bend right=20] (8)
 (4) edge[bend left=20] (2)
 (1) edge (6)
}

\begin{figure*}[!hbt]
\textit{1. Given the following network, where $\{3,4,7\}$ is a FVS and $\{3\}$ is a PFVS:}\\

\begin{tabular}{cc}
\raisebox{-1.5cm}{
\begin{tikzpicture}
\setlength{\unitlength}{1cm}
\coordinate (c1) at (120:\unitlength);
\coordinate (c2) at (40:1.4\unitlength);
\coordinate (c3) at (0,0);
\coordinate (c4) at (2\unitlength,0);
\coordinate (c5) at (-70:\unitlength);
\coordinate (c6) at (-130:\unitlength);
\coordinate[shift={(-\unitlength,0)}] (c7) at (c6);
\coordinate (c8) at (-35:2\unitlength);

\foreach \x in {1,2,5,6,8}{
	\node[vertex](\x) at (c\x) {\x};
}
\foreach \x in {3}{
	\node[vertex, fill=green](\x) at (c\x) {\x};
}
\foreach \x in {4,7}{
	\node[vertex, fill=yellow!75!red, text=white](\x) at (c\x) {\x};
}
\path[arcos,verde]\arcosPos;
\path[arcos,red]\arcosNeg;

\end{tikzpicture}} & \begin{tabular}{l}
$f_1 = \overline{x_3} \vee x_7$\\
$f_2 = \overline{x_4}$\\
$f_3 = (x_2 \wedge x_4) \vee (x_2 \wedge x_6) \vee (x_4 \wedge x_6)$\\
$f_4 = x_2 \vee \overline{x_8}$\\
$f_5 = x_3$\\
$f_6 = \overline{x_1} \vee x_5$\\
$f_7 = \overline{x_1} \vee x_8$\\
$f_8 = x_5 \wedge \overline{x_7}$
\end{tabular}
\end{tabular}\\

\textit{2. We choice an order compatible with the FVS and PFVS selected.}

\textit{In this case, we choice $\pi = (3,1,2,5,6,8,4,7).$}\\ 

\textit{3. If $a(3) = 0$, the execution of $(\funca)^{\pi}$ would be the following:}
\begin{center}
\begin{tabular}{|c|c|c|c|c|c|c|c|c|c|}
\hline
& & $x_1$ & $x_2$ & 	\cellcolor{green}$x_3$ & \cellcolor{yellow!75!red}$x_4$ & $x_5$ & $x_6$ & \cellcolor{yellow!75!red}$x_7$ & $x_8$\\
\hline
\multirow{8}{*}{$(\funca)^\pi$} & 3 & 0 & 0 & \highli{0} & 0 & 0 & 0 & 0 & 0 \\
& 1 & \highli{1} & 0 & \highli{0} & 0 & 0 & 0 & 0 & 0 \\
& 2 & \highli{1} & 1 & \highli{0} & 0 & 0 & 0 & 0 & 0 \\
& 5 & \highli{1} & 1 & \highli{0} & 0 & \highli{0} & 0 & 0 & 0 \\
& 6 & \highli{1} & 1 & \highli{0} & 0 & \highli{0} & \highli{0} & 0 & 0 \\
& 8 & \highli{1} & 1 & \highli{0} & 0 & \highli{0} & \highli{0} & 0 & \highli{0} \\
& 4 & \highli{1} & 1 & \highli{0} & \highli{1} & \highli{0} & \highli{0} & 0 & \highli{0} \\
& 7 & \highli{1} & 1 & \highli{0} & \highli{1} & \highli{0} & \highli{0} & \highli{0} & \highli{0} \\
\hline
\multirow{8}{*}{$((\funca)^\pi)^2$} & 3 & \highli{1} & 1 & \highli{0} & \highli{1} & \highli{0} & \highli{0} & \highli{0} & \highli{0} \\
& 1 & \highli{1} & 1 & \highli{0} & \highli{1} & \highli{0} & \highli{0} & \highli{0} & \highli{0} \\
& 2 & \highli{1} & \highli{0} & \highli{0} & \highli{1} & \highli{0} & \highli{0} & \highli{0} & \highli{0} \\
& 5 & \highli{1} & \highli{0} & \highli{0} & \highli{1} & \highli{0} & \highli{0} & \highli{0} & \highli{0} \\
& 6 & \highli{1} & \highli{0} & \highli{0} & \highli{1} & \highli{0} & \highli{0} & \highli{0} & \highli{0} \\
& 8 & \highli{1} & \highli{0} & \highli{0} & \highli{1} & \highli{0} & \highli{0} & \highli{0} & \highli{0} \\
& 4 & \highli{1} & \highli{0} & \highli{0} & \highli{1} & \highli{0} & \highli{0} & \highli{0} & \highli{0} \\
& 7 & \highli{1} & \highli{0} & \highli{0} & \highli{1} & \highli{0} & \highli{0} & \highli{0} & \highli{0} \\
\hline
\multirow{8}{*}{$((\funca)^\pi)^3$} & 3 & \highli{1} & \highli{0} & \highli{0} & \highli{1} & \highli{0} & \highli{0} & \highli{0} & \highli{0} \\
& 1 & \highli{1} & \highli{0} & \highli{0} & \highli{1} & \highli{0} & \highli{0} & \highli{0} & \highli{0} \\
& 2 & \highli{1} & \highli{0} & \highli{0} & \highli{1} & \highli{0} & \highli{0} & \highli{0} & \highli{0} \\
& 5 & \highli{1} & \highli{0} & \highli{0} & \highli{1} & \highli{0} & \highli{0} & \highli{0} & \highli{0} \\
& 6 & \highli{1} & \highli{0} & \highli{0} & \highli{1} & \highli{0} & \highli{0} & \highli{0} & \highli{0} \\
& 8 & \highli{1} & \highli{0} & \highli{0} & \highli{1} & \highli{0} & \highli{0} & \highli{0} & \highli{0} \\
& 4 & \highli{1} & \highli{0} & \highli{0} & \highli{1} & \highli{0} & \highli{0} & \highli{0} & \highli{0} \\
& 7 & \highli{1} & \highli{0} & \highli{0} & \highli{1} & \highli{0} & \highli{0} & \highli{0} & \highli{0} \\
\hline
\end{tabular}
\end{center}

\textit{4. Since $(\funca)(10010000) = 10010000$ and $f_3(1001000) = 0 = a(3)$, then $1001000$ is a fixed point of $f$}
\caption{Example of $(\funca)^{\pi}$ execution.}
\label{figA1}
\end{figure*}

\begin{figure*}
\textit{5. If $a(3) = 1$, the execution of $(\funca)^{\pi}$ would be the following:}
\begin{center}
\begin{tabular}{|c|c|c|c|c|c|c|c|c|c|}
\hline
& & $x_1$ & $x_2$ & 	\cellcolor{green}$x_3$ & \cellcolor{yellow!75!red}$x_4$ & $x_5$ & $x_6$ & \cellcolor{yellow!75!red}$x_7$ & $x_8$\\
\hline
\multirow{8}{*}{$(\funca)^\pi$} & 3 & 0 & 0 & \highli{1} & 0 & 0 & 0 & 0 & 0 \\
& 1 & 0 & 0 & \highli{1} & 0 & 0 & 0 & 0 & 0 \\
& 2 & 0 & 1 & \highli{1} & 0 & 0 & 0 & 0 & 0 \\
& 5 & 0 & 1 & \highli{1} & 0 & \highli{1} & 0 & 0 & 0 \\
& 6 & 0 & 1 & \highli{1} & 0 & \highli{1} & \highli{1} & 0 & 0 \\
& 8 & 0 & 1 & \highli{1} & 0 & \highli{1} & \highli{1} & 0 & 1 \\
& 4 & 0 & 1 & \highli{1} & 1 & \highli{1} & \highli{1} & 0 & 1 \\
& 7 & 0 & 1 & \highli{1} & 1 & \highli{1} & \highli{1} & 1 & 1 \\
\hline
\multirow{8}{*}{$((\funca)^\pi)^2$} & 3 & 0 & 1 & \highli{1} & 1 & \highli{1} & \highli{1} & 1 & 1 \\
& 1 & 1 & 1 & \highli{1} & 1 & \highli{1} & \highli{1} & 1 & 1 \\
& 2 & 1 & 0 & \highli{1} & 1 & \highli{1} & \highli{1} & 1 & 1 \\
& 5 & 1 & 0 & \highli{1} & 1 & \highli{1} & \highli{1} & 1 & 1 \\
& 6 & 1 & 0 & \highli{1} & 1 & \highli{1} & \highli{1} & 1 & 1 \\
& 8 & 1 & 0 & \highli{1} & 1 & \highli{1} & \highli{1} & 1 & 0 \\
& 4 & 1 & 0 & \highli{1} & 1 & \highli{1} & \highli{1} & 1 & 0 \\
& 7 & 1 & 0 & \highli{1} & 1 & \highli{1} & \highli{1} & 0 & 0 \\
\hline
\multirow{8}{*}{$((\funca)^\pi)^3$} & 3 & 1 & 0 & \highli{1} & 1 & \highli{1} & \highli{1} & 0 & 0 \\
& 1 & 0 & 0 & \highli{1} & 1 & \highli{1} & \highli{1} & 0 & 0 \\
& 2 & 0 & 0 & \highli{1} & 1 & \highli{1} & \highli{1} & 0 & 0 \\
& 5 & 0 & 0 & \highli{1} & 1 & \highli{1} & \highli{1} & 0 & 0 \\
& 6 & 0 & 0 & \highli{1} & 1 & \highli{1} & \highli{1} & 0 & 0 \\
& 8 & 0 & 0 & \highli{1} & 1 & \highli{1} & \highli{1} & 0 & 1 \\
& 4 & 0 & 0 & \highli{1} & 0 & \highli{1} & \highli{1} & 0 & 1 \\
& 7 & 0 & 0 & \highli{1} & 0 & \highli{1} & \highli{1} & 1 & 1 \\
\hline
\end{tabular}
\end{center}

\textit{6. Since $(\funca)(00101111) = 11101110$, does not exist fixed point of $f$ such that $f_3(x) = 1$}
\caption{Example of $(\funca)^{\pi}$ execution (continuation).}
\label{figA2}
\end{figure*}

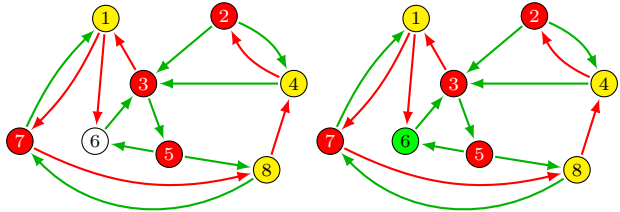
\begin{figure*}[!hbt]
\setlength{\unitlength}{1cm}
\begin{tabular}{p{4cm}c}
\textit{1. Given a labeled digraph and an order over its vertices (3,7,2,5,1,4,8,6). All vertices are added to $U$.} & \raisebox{-2cm}{\begin{tikzpicture}

\foreach \x in {1,...,8}{
	\node[vertex](\x) at (c\x) {\x};
}
\path[arcos,verde] \arcosPos;  
\path[arcos,red] \arcosNeg;
\end{tikzpicture}} \\
\textit{2. The vertex in $U$ with lower index (the vertex 3) is discarded from PFVS (added to $R$). Since $\forall v \in U \cup Y$, the subgraph induced by $R \cup \Set{v}$ has no circuits, a new vertex is selected.} & 
\raisebox{-3cm}{\begin{tikzpicture}
\foreach \x in {1,2,4,5,6,7,8}{
	\node[vertex](\x) at (c\x) {\x};
}
\foreach \x in {3}{
	\node[vertex, fill=red, text=white](\x) at (c\x) {\x};
}
\path[arcos,verde] \arcosPos;  
\path[arcos,red] \arcosNeg;
\end{tikzpicture}} \\
& \\
\textit{3. The next vertex (7) is discarded from PFVS. Since the subgraph induced by $R \cup \Set{1}$ has a negative circuit, (1) is added to $Y$. Similarly, (8) is added to $Y$.} & \raisebox{-2.75cm}{\begin{tikzpicture}
\foreach \x in {1,2,4,5,6,8}{
	\node[vertex](\x) at (c\x) {\x};
}
\foreach \x in {3,7}{
	\node[vertex, fill=red, text=white](\x) at (c\x) {\x};
}
\path[arcos,verde] \arcosPos;  
\path[arcos,red] \arcosNeg;
\end{tikzpicture}
\begin{tikzpicture}

\foreach \x in {2,4,5,6}{
	\node[vertex](\x) at (c\x) {\x};
}
\foreach \x in {3,7}{
	\node[vertex, fill=red, text=white](\x) at (c\x) {\x};
}
\foreach \x in {1,8}{
	\node[vertex, fill=yellow](\x) at (c\x) {\x};
}
\path[arcos,verde] \arcosPos;  
\path[arcos,red] \arcosNeg;
\end{tikzpicture}} \\
\textit{4. The next vertex (2) is discarded from PFVS. (4) is added to $Y$.} & \raisebox{-2cm}{\begin{tikzpicture}

\foreach \x in {4,5,6}{
	\node[vertex](\x) at (c\x) {\x};
}
\foreach \x in {3,7,2}{
	\node[vertex, fill=red, text=white](\x) at (c\x) {\x};
}
\foreach \x in {1,8}{
	\node[vertex, fill=yellow](\x) at (c\x) {\x};
}
\path[arcos,verde] \arcosPos;  \path[arcos,red] \arcosNeg;
\end{tikzpicture}
\begin{tikzpicture}
\foreach \x in {5,6}{
	\node[vertex](\x) at (c\x) {\x};
}
\foreach \x in {3,7,2}{
	\node[vertex, fill=red, text=white](\x) at (c\x) {\x};
}
\foreach \x in {1,8,4}{
	\node[vertex, fill=yellow](\x) at (c\x) {\x};
}
\path[arcos,verde] \arcosPos;  
\path[arcos,red] \arcosNeg;
\end{tikzpicture}} \\
\textit{5. The next vertex (5) is discarded from PFVS. Since the subgraph induced by $R \cup \Set{6}$ has a positive circuit, (6) is included in the PFVS (added to $G$). Since $U$ is empty, the first step is finished.} & \raisebox{-3cm}{\begin{tikzpicture}

\foreach \x in {6}{
	\node[vertex](\x) at (c\x) {\x};
}
\foreach \x in {3,7,2,5}{
	\node[vertex, fill=red, text=white](\x) at (c\x) {\x};
}
\foreach \x in {1,8,4}{
	\node[vertex, fill=yellow](\x) at (c\x) {\x};
}
\path[arcos,verde] \arcosPos;  
\path[arcos,red] \arcosNeg;
\end{tikzpicture}
\begin{tikzpicture}
\foreach \x in {6}{
	\node[vertex, fill=green](\x) at (c\x) {\x};
}
\foreach \x in {3,7,2,5}{
	\node[vertex, fill=red, text=white](\x) at (c\x) {\x};
}
\foreach \x in {1,8,4}{
	\node[vertex, fill=yellow](\x) at (c\x) {\x};
}
\path[arcos,verde] \arcosPos;  
\path[arcos,red] \arcosNeg;
\end{tikzpicture}} \\
\end{tabular}
\caption{Example of PFVS Algorithm (First step).}
\label{figB1}
\end{figure*}

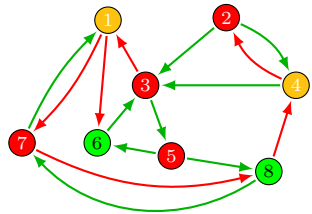
\begin{figure*}
\begin{tabular}{p{4cm}c}
\textit{6. For the second step (and following), all vertices in $Y$ are added to $U$.} & \raisebox{-2cm}{\begin{tikzpicture}
\foreach \x in {1,4,8}{
	\node[vertex](\x) at (c\x) {\x};
}
\foreach \x in {2,3,5,7}{
	\node[vertex, fill=red, text=white](\x) at (c\x) {\x};
}
\foreach \x in {6}{
	\node[vertex, fill=green](\x) at (c\x) {\x};
}
\path[arcos,verde] \arcosPos;  
\path[arcos,red] \arcosNeg;
\end{tikzpicture}} \\
\textit{7. The vertex in $U$ with lower index (1) is discarded from the PFVS (added to $R$ and $O$). Since the subgraph induced by $R \cup \Set{8}$ has a positive circuit (1,7,8,7,1), (8) is included in the PFVS.} & \raisebox{-3cm}{\begin{tikzpicture}

\foreach \x in {4,8}{
	\node[vertex](\x) at (c\x) {\x};
}
\foreach \x in {2,3,5,7}{
	\node[vertex, fill=red, text=white](\x) at (c\x) {\x};
}
\foreach \x in {6}{
	\node[vertex, fill=green](\x) at (c\x) {\x};
}
\foreach \x in {1}{
	\node[vertex, fill=yellow!75!red, text=white](\x) at (c\x) {\x};
}
\path[arcos,verde] \arcosPos;  
\path[arcos,red] \arcosNeg;
\end{tikzpicture}
\begin{tikzpicture}

\foreach \x in {4}{
	\node[vertex](\x) at (c\x) {\x};
}
\foreach \x in {2,3,5,7}{
	\node[vertex, fill=red, text=white](\x) at (c\x) {\x};
}
\foreach \x in {6,8}{
	\node[vertex, fill=green](\x) at (c\x) {\x};
}
\foreach \x in {1}{
	\node[vertex, fill=yellow!75!red, text=white](\x) at (c\x) {\x};
}
\path[arcos,verde] \arcosPos;  
\path[arcos,red] \arcosNeg;
\end{tikzpicture}} \\
& \\
\textit{8. The next vertex (4) is discarded from PFVS. Since $U$ is empty, the second step is finished. Since no vertex in $Y$, the algorithm ends.} & \raisebox{-2.7cm}{\begin{tikzpicture}

\foreach \x in {2,3,5,7}{
	\node[vertex, fill=red, text=white](\x) at (c\x) {\x};
}
\foreach \x in {6,8}{
	\node[vertex, fill=green](\x) at (c\x) {\x};
}
\foreach \x in {1,4}{
	\node[vertex, fill=yellow!75!red, text=white](\x) at (c\x) {\x};
}
\path[arcos,verde] \arcosPos;  
\path[arcos,red] \arcosNeg;
\end{tikzpicture}}
\end{tabular}
\caption{Example of PFVS Algorithm (Second step).}
\label{figB2}
\end{figure*}

\newpage
\begin{ejemplo}
\Cref{figB1,figB2} show the operation of the PFVS-Algorithm. Green vertices represent vertices in $P$, red vertices represent vertices in $R \setminus O$, yellow vertices represent vertices in $Y$, orange vertices represent vertices in $R$ and $O$ and white vertices represent vertices in $U$.
\end{ejemplo}

\begin{thebibliography}{99}

\bibitem{akutsu2009integer}
Tatsuya Akutsu, Morihiro Hayashida, and Takeyuki Tamura.
\newblock Integer programming-based methods for attractor detection and control
  of boolean networks.
\newblock In {\em Decision and Control, 2009 held jointly with the 2009 28th
  Chinese Control Conference. CDC/CCC 2009. Proceedings of the 48th IEEE
  Conference on}, pp. 5610--5617. IEEE, 2009.

\bibitem{akutsu1998identification}
Tatsuya Akutsu, Satoru Kuhara, Osamu Maruyama, and Satoru Miyano.
\newblock Identification of gene regulatory networks by strategic gene
  disruptions and gene overexpressions.
\newblock In {\em SODA}, volume~98, pp. 695--702, 1998.

\bibitem{akutsu1998system}
Tatsuya Akutsu, Satoru Kuhara, Osamu Maruyama, and Satoru Miyano.
\newblock A system for identifying genetic networks from gene expression
  patterns produced by gene disruptions and overexpressions.
\newblock {\em Genome Informatics}, 9:151--160, 1998.

\bibitem{akutsu2011determining}
Tatsuya Akutsu, Avraham~A Melkman, Takeyuki Tamura, and Masaki Yamamoto.
\newblock Determining a singleton attractor of a boolean network with nested
  canalyzing functions.
\newblock {\em Journal of Computational Biology}, 18(10):1275--1290, 2011.

\bibitem{albert2003topology}
R{\'e}ka Albert and Hans~G Othmer.
\newblock The topology of the regulatory interactions predicts the expression
  pattern of the segment polarity genes in drosophila melanogaster.
\newblock {\em Journal of theoretical biology}, 223(1):1--18, 2003.

\bibitem{Aracena08}
Julio Aracena.
\newblock Maximum number of fixed points in regulatory {B}oolean networks.
\newblock {\em Bulletin of Mathematical Biology}, 70(5):1398--1409, 2008.

\bibitem{aracena2018fixing}
Julio Aracena, Maximilien Gadouleau, Adrien Richard, and Lilian Salinas.
\newblock Fixing monotone boolean networks asynchronously.
\newblock {\em arXiv preprint arXiv:1802.02068}, 2018.

\bibitem{aracena2006regulatory}
Julio Aracena, Mauricio Gonz{\'a}lez, Alejandro Zu{\~n}iga, Marco~A Mendez, and
  Ver{\'o}nica Cambiazo.
\newblock Regulatory network for cell shape changes during drosophila ventral
  furrow formation.
\newblock {\em Journal of Theoretical Biology}, 239(1):49--62, 2006.

\bibitem{aracena2017fixed}
Julio Aracena, Adrien Richard, and Lilian Salinas.
\newblock Fixed points in conjunctive networks and maximal independent sets in
  graph contractions.
\newblock {\em Journal of Computer and System Sciences}, 88:145--163, 2017.

\bibitem{ARSsidma2017}
Julio Aracena, Adrien Richard, and Lilian Salinas.
\newblock Number of fixed points and disjoint cycles in monotone boolean
  networks.
\newblock {\em SIAM Journal on Discrete Mathematics}, 31(3):1702--1725, 2017.

\bibitem{bang2008digraphs}
J{\o}rgen Bang-Jensen and Gregory~Z Gutin.
\newblock {\em Digraphs: theory, algorithms and applications}.
\newblock Springer Science \& Business Media, 2008.

\bibitem{chen2008fixed}
Jianer Chen, Yang Liu, Songjian Lu, Barry O'sullivan, and Igor Razgon.
\newblock A fixed-parameter algorithm for the directed feedback vertex set
  problem.
\newblock {\em Journal of the ACM (JACM)}, 55(5):21, 2008.

\bibitem{davidich2008boolean}
Maria~I Davidich and Stefan Bornholdt.
\newblock Boolean network model predicts cell cycle sequence of fission yeast.
\newblock {\em PloS one}, 3(2):e1672, 2008.

\bibitem{devloo2003identification}
Vincent Devloo, Pierre Hansen, and Martine Labb{\'e}.
\newblock Identification of all steady states in large networks by logical
  analysis.
\newblock {\em Bulletin of mathematical biology}, 65(6):1025--1051, 2003.

\bibitem{dubrova2011sat}
Elena Dubrova and Maxim Teslenko.
\newblock A sat-based algorithm for finding attractors in synchronous boolean
  networks.
\newblock {\em IEEE/ACM transactions on computational biology and
  bioinformatics}, 8(5):1393--1399, 2011.

\bibitem{he2016algorithm}
Zhiwei He, Meng Zhan, Shuai Liu, Zebo Fang, and Chenggui Yao.
\newblock An algorithm for finding the singleton attractors and pre-images in
  strong-inhibition boolean networks.
\newblock {\em PloS one}, 11(11):e0166906, 2016.

\bibitem{helikar2013comprehensive}
Tom{\'a}{\v{s}} Helikar, Naomi Kochi, Bryan Kowal, Manjari Dimri, Mayumi
  Naramura, Srikumar~M Raja, Vimla Band, Hamid Band, and Jim~A Rogers.
\newblock A comprehensive, multi-scale dynamical model of erbb receptor signal
  transduction in human mammary epithelial cells.
\newblock {\em PloS one}, 8(4), 2013.

\bibitem{helikar2008emergent}
Tom{\'a}{\v{s}} Helikar, John Konvalina, Jack Heidel, and Jim~A Rogers.
\newblock Emergent decision-making in biological signal transduction networks.
\newblock {\em Proceedings of the National Academy of Sciences},
  105(6):1913--1918, 2008.

\bibitem{hinkelmann2011adam}
Franziska Hinkelmann, Madison Brandon, Bonny Guang, Rustin McNeill, Grigoriy
  Blekherman, Alan Veliz-Cuba, and Reinhard Laubenbacher.
\newblock Adam: analysis of discrete models of biological systems using
  computer algebra.
\newblock {\em BMC bioinformatics}, 12(1):295, 2011.

\bibitem{huang1999gene}
Sui Huang.
\newblock Gene expression profiling, genetic networks, and cellular states: an
  integrating concept for tumorigenesis and drug discovery.
\newblock {\em Journal of Molecular Medicine}, 77(6):469--480, 1999.

\bibitem{Karp}
Richard~M. Karp.
\newblock {\em Reducibility among Combinatorial Problems}, pp. 85--103.
\newblock Springer US, Boston, MA, 1972.

\bibitem{Kau69}
S.~A. Kauffman.
\newblock Metabolic stability and epigenesis in randomly connected nets.
\newblock {\em Journal of Theoretical Biology}, 22:437--467, 1969.

\bibitem{Kau93}
S.~A. Kauffman.
\newblock {\em Origins of Order Self-Organization and Selection in Evolution}.
\newblock Oxford University Press, 1993.

\bibitem{kazemzadeh2012boolean}
Laleh Kazemzadeh, Marija Cvijovic, and Dina Petranovic.
\newblock Boolean model of yeast apoptosis as a tool to study yeast and human
  apoptotic regulations.
\newblock {\em Frontiers in physiology}, 3:446, 2012.

\bibitem{klamt2006methodology}
Steffen Klamt, Julio Saez-Rodriguez, Jonathan~A Lindquist, Luca Simeoni, and
  Ernst~D Gilles.
\newblock A methodology for the structural and functional analysis of signaling
  and regulatory networks.
\newblock {\em BMC bioinformatics}, 7(1):56, 2006.

\bibitem{kobayashi2019design}
Koichi Kobayashi.
\newblock Design of fixed points in boolean networks using feedback vertex sets
  and model reduction.
\newblock {\em Complexity}, 2019, 2019.

\bibitem{li2004yeast}
Fangting Li, Tao Long, Ying Lu, Qi~Ouyang, and Chao Tang.
\newblock The yeast cell-cycle network is robustly designed.
\newblock {\em Proceedings of the National Academy of Sciences},
  101(14):4781--4786, 2004.

\bibitem{lovrics2014boolean}
Anna Lovrics, Yu~Gao, Bianka Juh{\'a}sz, Istv{\'a}n Bock, Helen~M Byrne,
  Andr{\'a}s Dinny{\'e}s, and Kriszti{\'a}n~A Kov{\'a}cs.
\newblock Boolean modelling reveals new regulatory connections between
  transcription factors orchestrating the development of the ventral spinal
  cord.
\newblock {\em PloS one}, 9(11):e111430, 2014.

\bibitem{madrahimov2013dynamics}
Alex Madrahimov, Tom{\'a}{\v{s}} Helikar, Bryan Kowal, Guoqing Lu, and Jim
  Rogers.
\newblock Dynamics of influenza virus and human host interactions during
  infection and replication cycle.
\newblock {\em Bulletin of mathematical biology}, 75(6):988--1011, 2013.

\bibitem{melkman2010determining}
Avraham~A Melkman, Takeyuki Tamura, and Tatsuya Akutsu.
\newblock Determining a singleton attractor of an and/or boolean network in o
  (1.587 n) time.
\newblock {\em Information Processing Letters}, 110(14-15):565--569, 2010.

\bibitem{mendoza2006method}
Luis Mendoza and Ioannis Xenarios.
\newblock A method for the generation of standardized qualitative dynamical
  systems of regulatory networks.
\newblock {\em Theoretical Biology and Medical Modelling}, 3(1):13, 2006.

\bibitem{montalva2008complexity}
Marco Montalva, Julio Aracena, and Anah{\'\i} Gajardo.
\newblock On the complexity of feedback set problems in signed digraphs.
\newblock {\em Electronic Notes in Discrete Mathematics}, 30:249--254, 2008.

\bibitem{raza2008logic}
Sobia Raza, Kevin~A Robertson, Paul~A Lacaze, David Page, Anton~J Enright,
  Peter Ghazal, and Tom~C Freeman.
\newblock A logic-based diagram of signalling pathways central to macrophage
  activation.
\newblock {\em BMC systems biology}, 2(1):36, 2008.

\bibitem{remy2008graphic}
{\'E}lisabeth Remy, Paul Ruet, and Denis Thieffry.
\newblock Graphic requirements for multistability and attractive cycles in a
  boolean dynamical framework.
\newblock {\em Advances in Applied Mathematics}, 41(3):335--350, 2008.

\bibitem{richard2018fixed}
Adrien Richard.
\newblock Fixed points and connections between positive and negative cycles in
  boolean networks.
\newblock {\em Discrete Applied Mathematics}, 243:1--10, 2018.

\bibitem{richard2019positive}
Adrien Richard.
\newblock Positive and negative cycles in boolean networks.
\newblock {\em Journal of theoretical biology}, 463:67--76, 2019.

\bibitem{Robert86}
F.~Robert.
\newblock {\em Discrete iterations: a metric study}, volume~6 of {\em Series in
  Computational Mathematics}.
\newblock Springer, 1986.

\bibitem{robert1995systemes}
Fran{\c{c}}ois Robert.
\newblock {\em Les systemes dynamiques discrets}, volume~19.
\newblock Springer Science \& Business Media, 1995.

\bibitem{robertson1999permanents}
Neil Robertson, Paul~D Seymour, and Robin Thomas.
\newblock Permanents, pfaffian orientations, and even directed circuits.
\newblock {\em Annals of Mathematics}, 150(3):929--975, 1999.

\bibitem{rolf2006improved}
Daniel Rolf.
\newblock Improved bound for the ppsz/sch{\"o}ning-algorithm for 3-sat.
\newblock {\em Journal on Satisfiability, Boolean Modeling and Computation},
  1:111--122, 2006.

\bibitem{saadatpour2011dynamical}
Assieh Saadatpour, Rui-Sheng Wang, Aijun Liao, Xin Liu, Thomas~P Loughran,
  Istv{\'a}n Albert, and R{\'e}ka Albert.
\newblock Dynamical and structural analysis of a t cell survival network
  identifies novel candidate therapeutic targets for large granular lymphocyte
  leukemia.
\newblock {\em PLoS computational biology}, 7(11), 2011.

\bibitem{saez2007logical}
Julio Saez-Rodriguez, Luca Simeoni, Jonathan~A Lindquist, Rebecca Hemenway,
  Ursula Bommhardt, Boerge Arndt, Utz-Uwe Haus, Robert Weismantel, Ernst~D
  Gilles, Steffen Klamt, et~al.
\newblock A logical model provides insights into t cell receptor signaling.
\newblock {\em PLoS computational biology}, 3(8), 2007.

\bibitem{samaga2009logic}
Regina Samaga, Julio Saez-Rodriguez, Leonidas~G Alexopoulos, Peter~K Sorger,
  and Steffen Klamt.
\newblock The logic of egfr/erbb signaling: theoretical properties and analysis
  of high-throughput data.
\newblock {\em PLoS computational biology}, 5(8), 2009.

\bibitem{sanchez2010arabidopsis}
Yara-Elena Sanchez-Corrales, Elena~R Alvarez-Buylla, and Luis Mendoza.
\newblock The arabidopsis thaliana flower organ specification gene regulatory
  network determines a robust differentiation process.
\newblock {\em Journal of theoretical biology}, 264(3):971--983, 2010.

\bibitem{singh2012boolean}
Amit Singh, Juliana~M Nascimento, Silke Kowar, Hauke Busch, and Melanie
  Boerries.
\newblock Boolean approach to signalling pathway modelling in hgf-induced
  keratinocyte migration.
\newblock {\em Bioinformatics}, 28(18):i495--i501, 2012.

\bibitem{tamura2009algorithms}
Takeyuki Tamura and Tatsuya Akutsu.
\newblock Algorithms for singleton attractor detection in planar and nonplanar
  and/or boolean networks.
\newblock {\em Mathematics in Computer Science}, 2(3):401--420, 2009.

\bibitem{tamura2009detecting}
Takeyuki Tamura and Tatsuya Akutsu.
\newblock Detecting a singleton attractor in a boolean network utilizing sat
  algorithms.
\newblock {\em IEICE Transactions on Fundamentals of Electronics,
  Communications and Computer Sciences}, 92(2):493--501, 2009.

\bibitem{Thom73}
Ren{\'e} Thomas.
\newblock Boolean formalization of genetic control circuits.
\newblock {\em Journal of Theoretical Biology}, 42(3):563--585, 1973.

\bibitem{Thom90}
Ren{\'e} Thomas and Richard d'Ari.
\newblock {\em Biological feedback}.
\newblock CRC press, 1990.

\bibitem{veliz2011reduction}
Alan Veliz-Cuba.
\newblock Reduction of boolean network models.
\newblock {\em Journal of theoretical biology}, 289:167--172, 2011.

\bibitem{veliz2014steady}
Alan Veliz-Cuba, Boris Aguilar, Franziska Hinkelmann, and Reinhard
  Laubenbacher.
\newblock Steady state analysis of boolean molecular network models via model
  reduction and computational algebra.
\newblock {\em BMC bioinformatics}, 15(1):221, 2014.

\bibitem{veliz2015dimension}
Alan Veliz-Cuba, Boris Aguilar, and Reinhard Laubenbacher.
\newblock Dimension reduction of large sparse and-not network models.
\newblock {\em Electronic Notes in Theoretical Computer Science}, 316:83--95,
  2015.

\bibitem{yang2013robustness}
Lijian Yang, Yan Meng, Chun Bao, Wangheng Liu, Chengzhang Ma, Anbang Li, Zhan
  Xuan, Ge~Shan, and Ya~Jia.
\newblock Robustness and backbone motif of a cancer network regulated by
  mir-17-92 cluster during the g1/s transition.
\newblock {\em PloS one}, 8(3):e57009, 2013.

\bibitem{zanudo2013effective}
Jorge~GT Za{\~n}udo and R{\'e}ka Albert.
\newblock An effective network reduction approach to find the dynamical
  repertoire of discrete dynamic networks.
\newblock {\em Chaos: An Interdisciplinary Journal of Nonlinear Science},
  23(2):025111, 2013.

\bibitem{zhang2008network}
Ranran Zhang, Mithun~Vinod Shah, Jun Yang, Susan~B Nyland, Xin Liu, Jong~K Yun,
  R{\'e}ka Albert, and Thomas~P Loughran.
\newblock Network model of survival signaling in large granular lymphocyte
  leukemia.
\newblock {\em Proceedings of the National Academy of Sciences},
  105(42):16308--16313, 2008.

\bibitem{zhang2007algorithms}
Shu-Qin Zhang, Morihiro Hayashida, Tatsuya Akutsu, Wai-Ki Ching, and Michael~K
  Ng.
\newblock Algorithms for finding small attractors in boolean networks.
\newblock {\em EURASIP Journal on Bioinformatics and Systems Biology},
  2007(1):20180, 2007.

\bibitem{zou2013algorithm}
Yi~Ming Zou.
\newblock An algorithm for detecting fixed points of boolean network.
\newblock In {\em Complex Medical Engineering (CME), 2013 ICME International
  Conference On}, pp. 670--673. IEEE, 2013.

\end{thebibliography}
\end{document}